\documentclass[12pt,preprint]{aastex}

\shorttitle{Koester et al.}
\shortauthors{The MaxBCG Cluster Finder}

\begin{document}

\newcommand{\nrtwo}{$N_{gals}^{r200}$}
\newcommand{\arxiv}{www.Arxiv.org}
\newcommand{\prep}{in prep.}

\title{MaxBCG: A Red Sequence Galaxy Cluster Finder}
\author{Benjamin P. Koester\altaffilmark{1},\email{bkoester@umich.edu} 
  Timothy A. McKay\altaffilmark{1},\email{tamckay@umich.edu} 
  James Annis\altaffilmark{2}, 
  Risa H. Wechsler\altaffilmark{3,4,5}, 
  August E. Evrard\altaffilmark{1},\\ 
  Eduardo Rozo\altaffilmark{3,6},  
  Lindsey Bleem\altaffilmark{6},
  Erin S. Sheldon\altaffilmark{7},  
  David Johnston\altaffilmark{8}
}

\altaffiltext{1}{Department of Physics, University of Michigan, Ann
  Arbor, MI 48104} 
\altaffiltext{2}{NASA/Fermilab Astrophysics Center,
  Fermi National Accelerator Laboratory, Batavia, IL}
\altaffiltext{3}{Kavli Institute for Cosmological Physics, The
  University of Chicago, Chicago, IL 60637}
\altaffiltext{4}{Department of Astronomy and Astrophysics and Enrico
  Fermi Institute, The University of Chicago, Chicago, IL 60637}
\altaffiltext{5}{Hubble Fellow} 
\altaffiltext{6}{Department of Physics, The University of Chicago, Chicago, IL 60637}
\altaffiltext{7}{Center for Cosmology and Particle Physics and
  Department of Physics, New York University, New York, NY 10003}
\altaffiltext{8}{Jet Propulsion Laboratory, Caltech, Pasadena, CA
  91109}

\begin{abstract}
Measurements of galaxy cluster abundances, clustering properties, and mass-to-light ratios in current and future surveys can provide important cosmological constraints. Digital wide-field imaging surveys, the recently-demonstrated fidelity of red-sequence cluster detection techniques, and a new generation of realistic mock galaxy surveys provide the means for construction of large, cosmologically-interesting cluster samples, whose selection and
properties can be understood in unprecedented depth. We present the details of the ``maxBCG'' algorithm, a cluster-detection technique tailored to multi-band CCD-imaging data. MaxBCG primarily relies on an observational cornerstone of massive galaxy clusters: they are marked by an overdensity of bright, uniformly red galaxies. This detection scheme also exploits classical brightest cluster galaxies (BCGs), which are often found at the center of these same massive clusters. We study the algorithm herein through its performance on large, realistic, mock galaxy catalogs, which reveal that it is $>90\%$ pure for clusters at $0.1 < z < 0.3$ with 10 or more red galaxies, and $>90\%$ complete for halos at $0.1 < z < 0.3$ with masses of $> 2 \times10^{14}h^{-1}M_{\sun}$. MaxBCG is able to approximately recover the underlying halo abundance function, and assign cluster richnesses strongly coupled to the underlying halo properties. The same tests indicate that maxBCG rarely fragments halos, occasionally overmerges line-of-sight neighboring ($\simeq10h^{-1}$ Mpc) halos, and overestimates the intrinsic halo red sequence galaxy population by no more than 20\%. This study concludes with a discussion of considerations for cosmological measurements with such catalogs, including modeling the selection function, the role of photometric errors, the possible cosmological dependence of richness measurements, and fair cluster selection across broad redshift ranges employing
multiple bandpasses.

\end{abstract}


\section{Introduction}
Clusters of galaxies are the largest bound objects in the
Universe. They are the likely observational counterparts of dark
matter halos, whose masses, abundance, and distribution are sensitive
probes of cosmology. Dark matter dominates their matter content, with
baryons forming no more than $\simeq 15\%$ of the total mass
\citep[and references therein]{voit05}. The baryonic component is
comprised of stars bound to constituent galaxies, stars that make up
the intra-cluster light \citep{zwicky51,gonzalez05,krick06}, and a hot
intra-cluster medium which accounts for $\simeq 7/8$ of the baryons
\citep{david95,evrard97,allen02}. In addition to fueling stars, the
gas component is responsible for X-ray emission \citep{sarazin86,
  rosati02} and Sunyaev-Zeldovich decrements \citep{carlstrom02}, all
of which make possible the baryonic detection of dark matter halos.

In cosmological measurements, the high end of the halo mass function
is the most observationally accessible. In this regime, cluster
abundances are most sensitive to changes in the proper distance at low
redshift. By $z \gtrsim 1$, abundances become more sensitive to the
growth function \citep{haiman01,levine02,battye03}. This sensitivity
has motivated numerous X-ray and optical searches for the most massive
clusters across a range of redshifts. Observations provide constraints
on the abundance of clusters as a function of observables such as
X-ray luminosity or optical richness. Extracting cosmological
constraints from these observations requires a good understanding of
the relationship between these observables and the underlying cluster
mass.

Among the various cluster detection schemes, optical imaging catalogs
enjoyed the earliest successes, targeting the high S/N end of the
abundance function across a range of redshifts. These began with the
early photographic plate surveys of Abell \citep{abell58,abell89} and
Zwicky \citep{zwicky68} and have since moved to digitized plate
catalogs \citep{lumsden92,dalton97,gal00} and CCD-imaging catalogs
\citep{postman96,annis99,olsen99,lobo00,goto02,gladders05}. During the
1980's, large spectroscopic catalogs were also generated, drawing from
a host of redshift surveys including the CfA redshift survey,
\citep[and many others]{huchra82}, the Nearby Galaxies Catalog
\citep{tully87}, the ESO Slice Project \citep{ramella99}, the Las
Campanas Redshift Survey \citep{tucker00}, the Nearby Optical Galaxy
Sample \citep{giuricin00}, the Southern Sky Redshift Survey
\citep{ramella02}, the 2df redshift survey
\citep{merchan02,eke04,yang05}, the Sloan Digital Sky Survey
\citep{miller05,berlind06}, and DEEP2 \citep{gerke05}.

At about the same time, space-based X-ray telescopes became an
invaluable tool for cluster science, detecting hundreds of X-ray
luminous clusters in flux limited surveys
\citep{schwartz78,gioia90,ebeling98,bohringer01,bohringer04}.
Individual clusters have been detected in this way out to $z\simeq
1.4$ \citep{mullis05,brodwin05}. More recently, the Sunyaev-Zeldovich Effect 
has been cited as a powerful means for cluster detection \citep{carlstrom02}. Imminent ground-based
Sunyaev-Zeldovich surveys promise nearly redshift-independent
detections of a range of halos \citep{ruhl04,loh05}.

While finding massive clusters has become more routine, returning
robust cosmological constraints has proven challenging.  For any
method of detecting clusters, robust cosmological constraints are
impossible without a reliable estimate of the absolute calibration of
and scatter in the observable--mass relation.  Scatter can be due to
both the stochastic nature of the physics that relates the observables
to mass, and to noise in the measurements themselves.  Because the
abundance function drops precipitously with increasing mass, any
attempted selection of dark matter halos imposed by a cut on an
observable in this high-mass regime scatters systems asymmetrically
above and below this cut, a selection effect known as Eddington bias
\citep{eddington13}.  In the exponential tail of the distribution,
the number of objects scattered up across the threshold in the 
observable considerably exceeds the number scattered down, and 
depending on the size of the scatter, this difference can be comparable 
to the sample size itself. \citep[e.g.,][]{lima05}.

The quality of the cosmological constraints from any measurement of
cluster abundances will depend on the amount of scatter between
the observable and halo mass, on the dynamic range of masses probed,
and on how well the mass--observable relation can be calibrated.
Thus interpretation of these abundance measurements
typically requires complex modeling strategies
\citep{ikebe02,holder00,levine02,battye03} or self-calibration
techniques \citep{majumdar03,majumdar04,lima05} to calibrate them and
to account for the selection biases incurred by the scatter in the
mass-observable relation.  In addition, self-calibration techniques
may be more challenging than previously expected if halo clustering
depends on properties other than mass \citep{wechsler06a}.

Despite their challenges, optically-selected cluster catalogs offer
many advantages over other methods.  Optical surveys are generally
able to detect individual galaxies at high S/N to high redshift and in
wide fields. This has enabled spectroscopic surveys to identify halos
at the group scale. Imaging catalogs can in principle do the same,
over a substantially larger redshift range, if they can limit the
degrading effects of projection. By contrast, X-ray surveys are
relatively insensitive to projection issues
\citep{ebeling98,bohringer00} and can be used to detect
intermediate-mass groups. Unfortunately the relatively low X-ray
luminosity of groups limits detection to the most luminous or low
redshift objects.

In addition to the obvious advantage of more bang for the buck if
photometric surveys can be used to identify clusters robustly ---
indeed, such surveys are often done primarily for other reasons, such
as the measurement of the galaxy power spectrum --- there are
substantial advantages to identifying systems farther down the
abundance function.  First, cosmological constraints are improved when
one can measure the mass function over a larger dynamic range: e.g.,
the {\em difference} in the abundance of low mass to high mass systems
can constrain the normalization of the power spectrum
with less degeneracy than the pure abundance of the highest mass
systems.  Second, the effect that scatter has on weakening constraints
are somewhat mitigated when one can identify systems farther down the
abundance function, where the mass function is shallower.

The other strong advantage of optical surveys over other methods is
the vast amount of additional data available on the clusters, that can
both contribute to mass calibration and that can inform our
understanding of galaxy evolution.  In addition to measurements of
cluster clustering \citep[e.g.][]{gonzalez02} that are possible with
other types of selection, the galaxies used to identify the clusters
can also be used for weak lensing measurements \citep{sheldon01,
  sheldon06}, and for velocity measurements if some or all of the
galaxies have spectra \citep[e.g.,][]{yang05, mckay06}. They can also
be used to measure the luminosity function and profiles of galaxy
clusters \citep{hansen05, popesso05}, and the cluster
mass-to-light-ratio \citep{tinker05, popesso06}, which can improve
cosmological constraints when combined with galaxy clustering
measurements \citep{vandenbosch03}.

As we will show, in imaging surveys, cluster catalogs that span a
broad range of masses and redshifts are a possibility. Minimizing
projection is the primary driver behind modern cluster and
group-finding algorithms. In recent years, imaging-based techniques
have matured with the construction of galaxy catalogs from surveys
such as the Digitized Second Palomar Observatory Sky Survey
\citep[DPOSS;][]{djorgovski99}, the Palomar Distant Cluster Survey
\citep[PDCS;][]{postman96}, the Red-Sequence Cluster Survey
\citep[RCS;][]{gladders05}, and the Sloan Digital Sky Survey
\citep[SDSS;][]{york00} to name a few. Automated efforts were
initiated in photographic plates, beginning with counts-in-cells
techniques \citep{couch91,lidman96} and adaptive kernel estimators
\citep{gal00,gal03}. Around the same time the powerful matched-filter
\citep{postman96} and adaptive matched filter \citep{kepner99} codes
were developed for use on the new CCD imaging. The adaptive kernel
method of \citet{gal00} was among the first to employ color cuts
conducive to cluster galaxy selection in their codes. The Red-Sequence
method \citep{gladders00} was the first to explicitly use color
selection of red-sequence galaxies as the primary lever arm for
cluster detection. The cut-and-enhance \citep{goto02} and the
spectroscopic C4 algorithms \citep{miller05} generalized this notion
of color uniformity to clustering in any color.

The guiding principles usually invoked in the appraisal of these
cluster finders are that the algorithm must be automated and
objective, should impose minimal constraints on cluster properties,
have a well-understood selection function, and should provide physical
properties of the clusters, such as redshift, luminosity, and richness
\citep{gal03,gal06}. The preceding algorithms address these points
with varying degrees of success, and primarily concern themselves with
the optically-richest systems. Since cosmology with clusters of
galaxies is an integral component of many current and future surveys,
cluster-finding algorithms that exploit the full power of modern
imaging surveys will become increasingly important as well. In
addition to uncovering the most massive halos, it is desirable that
they should also produce cluster catalogs that extend down the mass
function, to measure the mass function over the largest possible
dynamic range.

Evidence from studies of galaxy environments suggest a means by which
to accomplish this. Galaxies in dense environments preferentially
exhibit both early-type morphologies
\citep{hashimoto99,goto03,kuehn05} and red colors
\citep{balogh04,hogg04,tanaka04}. Indeed, \citet[][and references
  therein] {weinmann06} confirm that the red fraction in clusters
increases with increasing halo mass and luminosity, and decreasing
halo-centric radius. Thus, the presence of a nascent red galaxy
population in lower mass, lower richness halos, although less evolved
than the red galaxy population of massive clusters, provides the
opportunity to extend cluster detection to lower mass scales.

In this work, we present ``maxBCG'', an optical cluster-finding
algorithm adapted specifically to wide-field, multi-band digital
imaging surveys, such as the Sloan Digital Sky Survey (SDSS).  This
algorithm takes advantage of observational cornerstones of cluster
galaxies: they are typically the brightest galaxies at a given
redshift; their brightest members share very similar colors; and they
are spatially clustered. As we will show, these properties alone are
sufficient to robustly select and center the richest clusters, while
the power of the red-sequence takes over at lower richness to select
group-sized objects. Combining these factors in an algorithm, and
running it on surveys with the size and quality of the SDSS imaging
enables the generation of immense, high-quality cluster catalogs whose
selection function can be quantitatively determined. Between $0.1 < z
< 0.3$, maxBCG can reliably identify objects with $ > 10$ red
galaxies, and masses of $\sim 2 \times 10^{14}M_{\sun}$ in a volume
limited way.

Section \ref{sec:algo} of this paper contains an outline of the maxBCG
algorithm, and the details of its components. The execution of the
algorithm is covered in \S \ref{sec:likelihoods}, followed by study of
the selection function and performance of the algorithm in mock galaxy
catalogs in \S \ref{sec:selection}. A discussion and summary conclude
the paper. Throughout this paper, we assume $\Omega_m=0.3,
\Omega_{\Lambda}=0.7$ and $h=1$. A companion paper, \cite{koester06},
presents a catalog of 13,823 clusters identified using this method
from SDSS data, over the redshift range $0.1 < z < 0.3$.

\section{Algorithm}
\label{sec:algo}
\subsection{Outline}
We introduce the maxBCG algorithm, a new cluster detection technique
which exploits three primary features of galaxy clusters. The first is
the obvious spatial clustering of galaxies in clusters, which falls
off as $\sim 1/r$ projected in 2 dimensions (Section 2.3). The second is that the
most luminous cluster galaxies inhabit a tight sequence in the
color-magnitude diagram (CMD), the so-called ``E/S0 ridgeline''.  The
galaxies in the E/S0 ridgeline have very uniform colors, and are among
the reddest, brightest, and rarest galaxies at a given
redshift. Because of the strong 4000 \AA break in their rest-frame
spectra, their color is tightly correlated with redshift, so that
color measurements have the additional advantage that they provide
accurate redshift estimates.

The last feature is that there often exists a unique brightest cluster
galaxy (BCG) that resides in the E/S0 ridgeline, is typically
coincident with the center of the galaxy distribution, and is nearly
at rest relative to the halo center. BCGs in rich clusters usually
take the form of giant elliptical galaxies, so large that they are
only found at the centers of galaxy clusters. Thus they provide
important additional information about cluster locations and redshifts.

These features are folded into a likelihood function which is
redshift-dependent. Individual objects in an input photometric galaxy
catalog are evaluated at an array of redshifts with this composite
function, to assess the likelihood that they are BCGs living in an
overdense environment consisting of galaxies with a small dispersion
in color. An outline of algorithm is as follows:

1. Using the likelihood function, each object in an input galaxy
  catalog is tested at an array of redshifts for the likelihood that
  it is a cluster center.

2. Each object is assigned the redshift which maximizes this likelihood function.

3. The objects are ranked by these maximum likelihoods.

4. The object with the highest likelihood in the list becomes the
  first cluster center. All other objects within $z= \pm
  0.02$ (the typical $\sigma_z$ on a red galaxy), 
  a scaled radius $r_{200}$, and lower maximum likelihood are
  removed from the list of potential centers.

5, The next object in the list is handled similarly, and the
  process is continued, flagging other potential cluster centers
  within that object's neighborhood which have lower likelihoods.

6. All unflagged objects at the end of this percolation are kept,
  and are taken as BCGs identifying clusters in the final cluster
  list.

As a consequence of this method, each cluster gets a photometric
redshift, the redshift which maximizes its likelihood of being a
cluster center. During the process, the number of galaxies within
$1h^{-1}$ Mpc, within $\pm 2 \sigma$ of the E/S0 ridgeline, and
brighter than some minimum luminosity, $L_{min}$ (see section 2.5),
and dimmer than the BCG is recorded as $N_{gals}$, an initial richness
estimate. A scaled richness estimate, $N_{gals}^{r200}$, is also
generated. The only difference from $N_{gals}$ is that instead of
using a fixed aperture, it counts objects within $r_{200}$ of the BCG,
where $r_{200}=0.156N_{gals}^{0.6}h^{-1}$ Mpc \citep{hansen05}.

In the following, the components of the likelihood function encoding
these properties are outlined and the overall likelihood function is
described. The section concludes with a description of the input
galaxy catalog. Along the way we justify the choices made in design of
the algorithm.

\subsection{Likelihood Framework}
The general strategy for finding clusters in maxBCG is to maximize the
likelihood that a galaxy resides at the center of a cluster by varying
the redshift. This is a two-part likelihood function. One part checks
a galaxy for its similarity to a BCG, and the other part measures the
match of a galaxy's environment to the E/S0 ridgeline. In what
follows, we build the components of the likelihood:
\begin{equation}
\mathcal{L}_{tot}(z)=\mathcal{L}_{BCG}(z)\mathcal{L}_R(z)
\end{equation}
where $\mathcal{L}_{BCG}$ and $\mathcal{L}_{R}$ are the BCG and
ridgeline likelihoods.  The algorithm proceeds by evaluating this full
likelihood function for the whole galaxy density field and labeling
peaks in the field.

\subsection{Ridgeline Likelihood}
The ridgeline likelihood is broken up into spatial and color filters,
which are folded into a matched-filter likelihood. Each component
contains parameters that are driven observationally, such as the width
of the E/S0 ridgeline, or theoretically, such as the NFW density
\citep{navarro96} profile.
\subsubsection{Spatial Filter}
N-body simulations and studies of galaxy distributions in rich
clusters \citep[e.g., on early versions of this
  algortihm,][]{hansen05} have shown that both the distribution of
dark matter around dark matter halos and the distribution of galaxies
around cluster centers can be well-modeled by an NFW profile.  In
three dimensions,
\begin{equation}
\rho(r)=\delta_c\rho_c\frac{1}{(r/r_s)(1+r/r_s)^2},
\end{equation}
where $r_s=r_{200}/c$ is a scale radius, $c$ and $\delta_c$ are
dimensionless parameters, and $\rho_c$ is the critical density. With
$x=r/r_s$, \citet{bartelmann96} has shown that this can be written as a
surface density:
\begin{equation}
\Sigma(x)=\frac{2\rho_s r_s}{x^2-1}f(x)
\end{equation}
with 
%\begin{equation}
%f(x)=
%\begin{cases}
%1-\frac{2}{\sqrt{x^2-1}}\arctan{\sqrt{\frac{x-1}{x+1}}} & $(x>1)$ \cr
%1-\frac{2}{\sqrt{1-x^2}}\mathrm{arctanh}(\sqrt{\frac{1-x}{x+1}}) & $(x<1)$ \cr
%0 & \text{($x=1$)}
%\cr 0 & \text{($x>20$)}
%\end{cases}
%\end{equation}
\begin{equation}
  f(x) = \cases{
    1-{2 \over \sqrt{x^2-1}}\ \mbox{tan}^{-1} \sqrt{{x-1\over x+1}} & $x>1$ \cr
    1-{2 \over \sqrt{1-x^2}}\ \mbox{tanh}^{-1} \sqrt{{1-x\over x+1}} & $x<1$ \cr
    0 & $x=1$ \cr
    0 & $x>20$
  }
\end{equation}

Because $f(x)$ diverges at small $x$, and because this assumption is
shakier at small radii, the profile is truncated at $x=0.667$, or
$r=100 h^{-1}$ kpc (see below). Objects near the center are given
large weights, and those far away are strongly down-weighted. This
model allows for a variable scale radius, $r_s$ that could be
additionally maximized to provide an optimal size for each cluster. We
choose a fixed $r_s=150$ kpc in this implementation. This quantity is
normalized to 1 by integrating the area over $0.0 < r < 3h^{-1}$ Mpc
($0.0 < x < 20$). Tests show that
varying $r_s$ between 100 kpc and $1h^{-1}$ Mpc only has a very
marginal influence on the overall abundance function or the algorithm
selection function. In general, and in accordance with other authors
\citep{lubin96,gladders00}, we find that the exact parameters
of the radial likelihood function have only a minimal effect on the
performance. Only extremes, such as top-hat spatial filters that have
sharp edges, strongly influence the performance.

\subsubsection{Color Filter}
Investigations of rich clusters indicate the presence of a universal
red-sequence, extending from the nearby Coma and Virgo clusters
\citep{bower92}, through intermediate redshifts
\citep{smail98,barrientos99}, and out to redshifts as high as $z \sim
1.4$ \citep{mullis05,eisenhardt05}. This population of galaxies
dominates the bright end of the cluster luminosity function
\citep{sandage85,barger98} and consists of E and S0 galaxies objects
with a narrow scatter in color, hence its designation as the ``E/S0
ridgeline'' \citep[Figure 1]{visvanathan77,annis99}. This tight sequence is
due to the presence of the uniformly old underlying stellar
populations in these galaxies, resulting from passive evolution and
minimal star formation. In Coma and Virgo, the stellar component has
most likely been in place for at least 2 Gyr \citep{bower92}. For a
review of the red sequence in galaxy clusters, see \citet{gladders00}
and references therein.

The red sequence is not limited to galaxy clusters. It persists
smoothly down to the lower density environments harbored in groups
\citep{postman84,zabludoff98,tran01}. Transformation of a field spiral
into a red-sequence elliptical has been hypothesized to occur in two
steps \citep[e.g.,][]{weinmann06}: 1) mergers between spiral galaxies
in low velocity groups create ellipticals \citep{toomre72}, which in
turn 2) are stripped of their hot gas through processes such as ram
pressure stripping and galaxy harassment, effectively truncating star
formation \citep{larson80,balogh00}. This picture has been
supplemented by recent simulations which implicate active galactic
nucleus (AGN) feedback in truncating star-formation
\citep[e.g.,][]{croton05}. Depending on the state of the evolution of
the group and the possible presence of AGN, the E/S0 ridgeline may be
ill-defined in a given group. However, the presence of red-sequence
galaxies is ubiquitous in higher mass groups, and extends
substantially farther down the richness function than any properties
that depend on cluster gas properties.

Motivated by the location of the 4000 \AA\ break and the luminous red
galaxy (LRG) cuts outlined by \citet{eisenstein01}, we employ cuts in
SDSS $g-r$ and $r-i$ in our search (Section 3), which depend on the color-redshift 
relation of the E/S0 ridgeline. First, we measured the width of the ridgeline using stacked Abell clusters at $z\simeq 0.1$ in the SDSS spectroscopic survey,
and fit a line to the color-magnitude diagrams, as there is a small tilt. We
then projected the distribution along the line. The resulting color distributions are well-fit by Gaussians with widths if 0.05 and 0.06 in $g-r$ and $r-i$,
in agreement with a range of studies
\citep{bower92,smail98,lopezcruz04,barrientos04}; since the color errors are small ($\simeq 0.005$ mag) for this bright sample ($r \lesssim 17.7$), most of the width is intrinsic to the ridgeline. This information is
folded into $\mathcal{L}_{R}$ for some color $j-k$ using a normalized
function $G_{j-k}$, of the following form:
\begin{equation}
G_{j-k}(z)=\frac{1}{\sqrt{2\pi}\sigma}\exp\frac{(x_{j-k}-\bar{x}(z))^2}{2\sigma^2}
\end{equation}
where $x_{j-k}$ is the color of some galaxy being tested, $\bar{x}(z)$
is the predicted E/S0 ridgeline color at some redshift (see below) and
the width, $\sigma$, is determined according to
\begin{equation}
\sigma=\sqrt{\sigma_{j-k}^2+(\sigma_{j-k}^r)^2}
\label{eq:ridgelinewidth}
\end{equation}
Here, $\sigma_{j-k}$ is the error in the measured color of an
individual galaxy and $\sigma_{j-k}^r$ is the intrinsic width of the
E/S0 ridgeline, 0.05 for $g-r$ and 0.06 for $r-i$. The ridgeline width
is taken to be constant with both redshift and richness, a reasonable
approximation for the redshift range considered here. It is clear that
this function will peak in the field of a rich cluster when the right
redshift-color combination is tested, and that the peak will be strong
for galaxies with well-measured colors and for clusters with
especially tight E/S0 ridgelines.

One feature of this prescription is that the photometric error, $\sigma_{j-k}$, is folded into the cluster detection process. Galaxies with poorer color measurements are able to contribute to ridgelines at a wider range of redshifts, albeit at a suppressed level because of the breadth of $G_{j-k}$. Those with good color measurements have a very narrow, highly peaked $G_{j-k}$, such that they make a very strong contribution to the ridgeline at the right redshift, and a vanishingly small contribution to ridgelines at other nearby redshifts. The input of photometric errors to cluster detection is somewhat similar to that in \citet{gladders00}, where the color slices are defined according to the typical photometric error and the intrinisic ridgeline width. For comparison, \citet{goto02} use photometric error estimates to exclude galaxies from the cluster detection process whose color errors are larger than the expected ridgeline width. \citet{kim02} and \citet{gal00} do not use the photometric error explicitly in cluster detection.

As noted by \citet{gladders00}, the fiducial color--redshift model,
$\bar{x}(z)$, does not need to be perfect to find clusters; clustering
in color requires little knowledge of redshift. However, accurate
color--redshift relations are certainly helpful in providing accurate
photometric redshifts. This is a very reasonable demand, as the strong
4000\AA\ break feature in red-sequence galaxies makes $g-r$ an
effective indicator of redshift. This feature enables red-sequence
algorithms to deliver accurate photometric redshifts without
spectra. The accuracy of the photometric redshifts is borne out in the
accompanying maxBCG catalog paper \citep{koester06}, where the
photometric redshift errors are shown to be $\sigma_z <0.015$ for
clusters with \nrtwo\ $> 10$ and with $0.1 < z < 0.3$. For comparison, in the SDSS, the method of \citet{goto02} also returns photometric redshift errors of $\sigma_z =0.0147$ over a similar redshift range, while the hybrid-matched filter of \citet{kim02} has errors that range from $\sigma_z=0.007$ to 0.02. Outside the SDSS, the digitized NSC \citep{gal03} survey reaches $\sigma_z$ as low as 0.033, while the RCS \citep{gladders05} estimate $\sigma_z=0.05$ over $0.2 < z < 1.0$. 

In maxBCG, the determination of $\bar{x}(z)$ is driven
observationally, by the SDSS LRGs \citep{eisenstein01} whose colors
and redshifts serve as a template. An advantage of using the LRGs is
the fact that they provide observationally-determined colors and
redshifts of the early-type galaxies that we expect to be in dense
environments. A drawback is that the standard LRG color--magnitude
selection criteria are only valid for $z > 0.15$, so that LRGs outside
this range must be selected by alternate means. We handle this
shortcoming by using a combination of spectroscopic classification
(${\tt eclass < 0}$) and morphology (${\tt fracDev_{r}>0.8}$)
\citep{stoughton02,bernardi05} to extract luminous early-type galaxies
below $z=0.15$ from the SDSS full spectroscopic sample
\citep{adelman06}. The $g-r$ and $r-i$ colors of these
spectroscopically-identified objects were used to predict the
ridgeline colors of $z < 0.15$ clusters. A fiducial version of the
cluster finder was then run, and spectroscopically-identified cluster
members were used to create a refined prediction of the color--redshift
relation at $z < 0.15$.

Beyond $z=0.15$, where the standard LRG selection is robust, and
clusters with multiple member spectra are less common, the LRG colors
were used to select the initial cluster galaxies. Their colors
were then used to calibrate the color--redshift relation. The resulting
piecewise-defined function for determination of the ridgeline $g-r$
color vs. redshift is shown in Figure 2. A similar
relation exists for $r-i$. The combined $g-r$ and $r-i$ models
constructed in this way are at the heart of the ridgeline component of
the likelihood function, as well as the photometric redshifts provided
by this algorithm.

For comparison, the LRG $g-r$ colors are shown in Figure 2,
   as shown in Table 1 of \citet{eisenstein01}.
   The upper and lower short dashed lines are the
   g-r colors predicted from their Pegase simulations for passive
   (upper) and star-forming (lower) dotted lines. The LRG cuts in the SDSS
   are designed to select the brightest $\gtrsim 3 L_*$ red galaxies,
   which should also be the reddest. The upper short-dashed line indeed
   falls on the red end of the cluster galaxy colors at each redshift,
   and the star-forming model forms a sort of lower bound for the cluster
   members, so the cluster colors (and our model) compare quite well
   with the LRG sample.

\subsubsection{Constructing the Ridgeline Likelihood}
The general framework for developing $\mathcal{L}_{R}$ is identical to
that derived by \citet{postman96}. The following reviews this
process. The application of this formalism to our filters is valid as
long as the assumptions about the Gaussian character of the galaxy
number counts in a given angular aperture hold. Even when this begins
to fail, the efficacy of the red sequence and the fact that the
likelihoods are used as a ranking tools (\S \ref{sec:likelihoods}) and not as
absolute assignments of cluster significance allows us to use this
method of identification for relatively low mass groups.

The framework is constructed by first writing down the cluster
likelihood:
\begin{equation}
\mathcal{L}_{R} \sim \frac{1}{\sigma}\exp{\frac{[b(c)+\Lambda_N
      M(r,c)-D(r,c)]^2}{\sigma^2}}
\label{eq:clustlike}
\end{equation}
where $M(r,c)$ is the model for the cluster spatial and color
distribution and $\Lambda_N$ is a measure of the cluster richness. In
this work, we take $M(r,c)=\Sigma(r/r_s)G_{g-r}G_{r-i}$, described in
equations (3) and (5). This is essentially the contribution of a
galaxy with colors $c$, at a physical radius $r$ to the likelihood
that a test point is in a cluster ridgeline. The uncertainty,
$\sigma^2$, is Poisson since the number counts in the aperture are
dominated by background, $b(c)$. The Poisson nature then implies
$\sigma^2 \simeq b(c) $. We note that the Gaussian approximation is
assumed in Eq. \ref{eq:clustlike}; since the counts are essentially Poisson distributed
with a high mean, the Gaussian approximation is sufficient.

We have on hand a survey containing galaxies with colors and
magnitudes, $c$ and $m$, and radial distances $r$ from a point of
interest. Sitting at some point in the survey, the likelihood is
evaluated over the full survey, $D(r,c)$, such that:

%\begin{align}
%\ln\mathcal{L}_{R} \propto & \int_{Area,c}\ln{{\sigma}} \nonumber\\
%&+\int_{Area,c}{\frac{[b(c)-D(r,c)]^2}{\sigma^2}} \nonumber\\
%&+\int_{Area,c}{\frac{[\Sigma(r/r_s)G_{g-r}G_{r-i}]^2}{\sigma^2}}
%\end{align}

\begin{equation}
\ln\mathcal{L}_{R} \propto \int_{Area,c} \left( \ln{{\sigma}}+{\frac{[b(c)-D(r,c)]^2}{\sigma^2}} +{\frac{[\Sigma(r/r_s)G_{g-r}G_{r-i}]^2}{\sigma^2}} \right)
\end{equation}

It is apparent in the previous sections that the relevant quantities
in the likelihood are all functions of redshift. The models for the
colors and the angular scale of the spatial filter will be set after a 
redshift is chosen for likelihood evaluation (\S \ref{sec:likelihoods}). Given
that $\sigma^2 \simeq b(c)$, \citet{postman96} have shown that
maximizing this quantity is nearly the same as maximizing
\begin{equation}
\ln\mathcal{L}_{R} \propto \int{\frac{\Sigma(r/r_s)G_{g-r}G_{r-i}}{b(c)}\mathrm{d}^2r\mathrm{d}m\mathrm{d}c}
\end{equation}

The integral is evaluated by taking the input galaxies to be
$\delta$-functions at the radii and magnitudes at which they are
observed. This turns the integral into a sum over all the galaxies in
the survey (Eq. \ref{eq:sumovergals}). This likelihood equation is then applied in
following manner: Sit on some galaxy in the survey, a ``potential''
BCG (section 2.4) at a location $\theta$.  Look at each neighboring
galaxy, $k$, of $N_{g}$ total neighboring galaxies. Each neighboring
galaxy has colors $c_k$, and projected distance $r_k$ from the
potential BCG. Compute the product $\Sigma(r/r_s)G_{g-r}G_{r-i}$ for
each neighboring galaxy, and sum:
\begin{equation}
S(\theta)=\sum^{N_{g}}_{k=1} \Sigma [r_k(\theta)]G_{g-r}(c_k)G_{r-i}(c_k)
\label{eq:sumovergals}
\end{equation}
The sum $S(\theta)$ consequently embodies the measure of a galaxy's
environment, and serves as the ridgeline likelihood.

\subsection{BCG Likelihood}
\subsubsection{Brightest Cluster Galaxies}
The second component of the full maxBCG likelihood is
$\mathcal{L}_{BCG}$, the likelihood that a galaxy is a brightest
cluster galaxy. By definition, every cluster has a BCG, and it usually
takes the form of a luminous early-type galaxy with an $r^{-1/4}$
surface brightness profile. In some cases the BCG is a giant ``cD''
galaxy, whose surface brightness is shallower than $r^{-1/4}$ at large
radii, forming the cD-envelope. Their luminosities can reach $\sim
10L_*$ and the envelopes can extend up to $1$ Mpc
\citep{hoessel85}. Particularly in the richest clusters, these
galaxies comprise a statistically-distinct population
\citep{hansen05,loh06}. In measuring the cluster luminosity function,
\citet{hansen05} have shown that excluding the BCG markedly improves the fit
to a Schechter function, and that the BCG luminosity function is
approximately Gaussian. The contrast between BCGs member galaxies for
poorer systems is less evident.

Because of the rarity of clusters, galaxies with BCG properties are
themselves rare. Using this extra information increases the fidelity
of cluster finding. With this likelihood, we look for individual
galaxies with specific properties, and fold this in with a measurement
of their environment as specified by the ridgeline likelihood. As in \citet{loh06}, 
who statistically selected BCGs from the SDSS LRG
sample on the basis of local density and magnitude, we find that
average BCGs make up the bright tip of the red-sequence. Their
location in the E/S0 ridgeline is clear in ridgeline plots in Figure 1.

In an effort to study the ubiquity of BCGs, and quantify the characteristics of BCGs, the combined NORAS-REFLEX sample described in \citet{koester06} was used to conduct a preliminary visual inspection of the 99 clusters from this combined catalog. We find that 79/99 ($\simeq 80\%$) of the clusters in this sample exhibit a single distinct BCG. Of the remaining 20 clusters, 15 display 2 BCG-like galaxies, and the remaining 5 have no clear BCG and are in fact rather optically-poor. A handful of those with 2 BCG-like galaxies appear as if they are two separate systems undergoing mergers. Of the 79 distinct BCGs, 74 are within $\pm 0.1$ in $g-r$ of the ridgeline for the cluster ($\simeq 94\%$), and 82 out of the full sample of 99 fell within the ridgeline ($\simeq 83\%$).

The BCGs in these clusters fall very near the X--ray centers as well. The median separation of the 79 distinct, visually-identified BCGs, is $54 h^{-1}$ kpc. When we add in the clusters with 2 BCGs, the median separation rises to $77 h^{-1}$ kpc, and the median separation of the full 99 clusters is $220 h^{-1}$ kpc. 

The reader is also referred to \citet{koester06}, where it is revealed that in the final cluster catalog, the maxBCG algorithm centers clusters within a median $57 h^{-1}$ kpc of the X--ray center for this same NORAS-REFLEX sample.

Taken together, these results demonstrate that classical BCGs fall within the E/S0 ridgeline better than $90\%$ of the time, and that they are quite near the center of the cluster, both of which are encouraging for the task at hand. Nevertheless, the complex nature of the cluster environment creates exceptions to this pattern, and this must be kept in mind as $\mathcal{L}_{BCG}$ is folded in the maxBCG technique.

\subsubsection{BCG Model}
$\mathcal{L}_{BCG}$ is calculated independently of environment. Our
goal is to tune maxBCG to find rich clusters, and to center those
clusters on the cD-like BCGs. Thus, we need to build a sizable sample
of training BCGs, enough to generate a model for the colors and
magnitudes of such objects. This observationally-driven model is
constructed in two steps. First, fiducial BCGs are chosen by
constructing a template consisting of bright ($M_r < -21$) LRGs and
their distribution in $g-r$, $r-i$, $i$, and $z$ space modeled with a
linear combination of Gaussians fit by expectation maximization
algorithm \citep{connolly00}. Below $z=0.15$, the SDSS spectroscopic
sample was used to select LRGs similar to ridgeline calibration in
\S 2.3. The maxBCG cluster finder is then run with these bright
LRGs serving as the BCG model for $\mathcal{L}_{BCG}$. Then, to
construct a refined BCG model, a set of visually identified BCGs from
the richest clusters across $0.1 < z <0.3$ are chosen from this first
run. We identify 100 BCGs in rich clusters with a characteristic cD
envelope, and a single nucleus. Because of their luminosities, most
have spectroscopic redshifts in the SDSS, so that it is possible to
track their colors and magnitudes as function of redshift.

The $r$-band magnitudes are plotted as a function of redshift for
these BCGs in Figure 3. In this figure, note that we plot
the $r$-band for comparison to the study of \citet{loh06}, but in
practice we use the $i$-band magnitude (see \S 2.5 and the
Discussion). There is a clear trend of magnitude with spectroscopic
redshift. We fit a quadratic function to this relation, and use it as
our model for the trend of BCG magnitude with redshift. The slope of
this magnitude--redshift relation is nearly identical to that in
\citet{loh06}, with an offset to brighter magnitudes, as expected. The
BCGs from rich objects in the full maxBCG catalog in \citep{koester06}
(greyscale, Figure 3) display a magnitude--redshift
relation similar in shape to that found by \citet{loh06}.

The BCG colors evolve with redshift in essentially the same way as the
ridgeline does (see Figure 1 for examples of BCGs in
the color--magnitude relation). This is consistent with the results
\citet{weinmann06}, who show that on average, central galaxies
properties correlate well with their satellite properties.

The likelihood, $\mathcal{L}_{BCG}$, is specified to be
\begin{equation}
\mathcal{L}_{R}(z)=G_{g-r}^{BCG}(z)G_{r-i}^{BCG}(z)\mathrm{e}^{-((m-m_i)/\sigma_c)^2}.
\end{equation}
As mentioned above, the Gaussians $G_{j-k}^{BCG}$ are nearly identical
to their ridgeline-likelihood counterparts. A crucial component of
this likelihood is the smooth cutoff function that ensures that a
galaxy's luminosity (indicated by its $i$-band magnitude) is high
enough to be a BCG at a given redshift. The cutoff, $m_i$, is taken
from the $i$-band magnitude--redshift relation, similar to that for the
$r$-band in Figure 3. The width $\sigma_c$, is taken from
\citet{loh06} to be 0.3 magnitudes. Fitting in this way to rich
clusters makes us most sensitive to the richest clusters that harbor
these BCGs, and further boosts their cluster likelihoods relative to
poorer systems. Through visual inspection, we find that this component
of the maxBCG likelihood is most effective in properly centering
clusters on what one would manually classify as the BCG.

\subsection{Input Galaxy Catalog}

For ease of processing, the input catalog is broken up into
redshift-dependent slices. In each slice we include galaxies whose
colors are within $3\sigma$ of the predicted $g-r$ and $r-i$ colors. A
redshift-dependent lower magnitude limit is also imposed in each
slice, such that galaxies with luminosities down to some $L_{min}$ are
measured for each cluster, regardless of redshift. This requirement
reflects the wish to measure richnesses and find clusters in a manner
that does not dependent on survey flux limits.

We define the cutoff magnitude, $a(z) + M_*$, which is designed to
consistently select objects above $M_{min}$ at all redshifts we
consider. $M_*$ is drawn from the galaxy luminosity function in
clusters, and $a(z)$ embodies distance, $k$-correction, and evolution of
$M*$. This cutoff is selected so that at the redshift limit of the
catalog, objects of brightness $a(z) + M_*$ have small photometric
errors and are still well within the ridgeline of typical clusters. We
work in the $i$-band when applying these cuts, as the $4000$\AA\ break
at $z<0.3$ stays well out of the range of the $i$-band, which is
centered at $\simeq 7400$\AA\ . The typical early-type spectra red-ward
of $5000$\AA\ are relatively flat compared to the break, so this
minimizes differences in luminosity estimates that would be due to
large spectral features moving into and out of bands.

We will shortly determine $a(z)$, but first it is instructive to
consider the relevant magnitude range we wish to target. $M^r_*$ in
the $r$-band for the galaxy population is -20.44 \citep{blanton03}, at
$z\simeq 0.1$, and rises to -20.75 in rich clusters
\citep{hansen05}. Using Table 1 in \citet{eisenstein01} for the
non-star-forming galaxy model, we can convert this characteristic
$r$-band magnitude for rich clusters to an approximate $i$-band
magnitude, $M^i_*$, at $z=0.1$. For this combination, $M^i_*=-21.24$,
which implies a luminosity $L_*=2.3 \times 10^{10}h^{-1}L{\sun}$. For
the galaxy catalog we use in \citet{koester06}, a cutoff of $0.4L_*$
is chosen, for which the maximum magnitude of the input galaxy
catalogs we have on hand, corresponds to a redshift limit of $\sim
0.40$. For the rest of this paper, we refer to this $0.4L_*$ as
``$L_{min}$''.

To actually determine the appropriate $k$-corrected magnitude cut at
each redshift range of interest, we use a Pegase-2 stellar
population/galaxy formation model, similar to that of
\citet{eisenstein01}. Briefly, a range of plausible scenarios was run
until the color distribution was very near that of the LRG/BCG
colors. A very metal-rich non-primordial gas model was chosen, and
predictions for the $i$-band magnitude of an $L_{min}$ galaxy over the
range of redshifts was output. The characteristic magnitudes and
luminosities are set at $M^i_*=-21.22$ in this model, which
corresponds to $2.25 \times 10^{10} L_{\sun}$. This compares well with
the value of $M^i_*=-21.24$ derived from the observed cluster
luminosity function. The corresponding $L_{min}$, as defined above, is
then $0.9 \times 10^{10}h^{-1}L{\sun}$, with an absolute magnitude of
$-20.25$ in the $i$-band. These outputs are saved as our
redshift-dependent magnitude limit $a(z) + M_*$.

One final magnitude-dependent step is enforced when actually running
the algorithm. In evaluating $\mathcal{L}_{R}$ to test some galaxy for
its similarity to a BCG, only galaxies with apparent magnitudes dimmer
than that of the potential BCG are included in $\mathcal{L}_{R}$. This
works in concert with $\mathcal{L}_{BCG}$ to select clear BCGs more
reliably; if there are two bright galaxies with otherwise nearly equal
likelihoods, the dimmer one will almost always receive the lower
ridgeline likelihood, by virtue of the fact that $\mathcal{L}_{R}$
increases with the number of galaxies evaluated. It also reduces the
number of bright foreground galaxies evaluated in the likelihood
function or included in the cluster membership. This cut only depends
on the magnitude of the object being tested as a BCG, and not the
redshift.

\section{Evaluating Likelihoods}
\label{sec:likelihoods}

As an alternative to pixelization of the galaxy catalog, every galaxy
in the survey is tested as a potential cluster center, and the
likelihoods are computed considering the possibility that any other
galaxy could be a member galaxy of the potential cluster. To speed up
computation, the following exceptions are made (in the following,
galaxies being evaluated as BCGs at clusters centers are called
``candidate BCGs'' and objects evaluated in the cluster likelihood are called
``neighbors''):

1. To be considered as a center, the candidate BCG must lie within $\pm 3
  \sigma$ in $g-r$ and $r-i$ of the predicted ridgeline colors at the
  assumed redshift, and brighter than $L_{min}$.

2. A neighbor galaxy must be within projected 3 $h^{-1}$ Mpc of the
  center.

3. A neighbor galaxy must be within $\pm 3 \sigma$ in $g-r$ and $r-i$
  at the test redshift, brighter than $L_{min}$, and dimmer than the
  candidate BCG.

In all these cuts, $\sigma$ is given by Eq. \ref{eq:ridgelinewidth}
for the appropriate color. The goal is to eliminate objects whose
colors, magnitudes, and angular separations are clearly inconsistent
with being red-sequence galaxies at the test redshift. A consequence
of item (1) is that each candidate BCG is tested at a range in redshift of $\sim
z \pm 0.05$, which we find is adequate to map out the maximum in the
likelihood. Figure 4 shows the shape of these
likelihood functions vs. redshift for a previously unknown maxBCG
cluster at $z=0.23$. In Figure 5, the SDSS image (see online Journal for color version) is
shown, centered on a bright BCG with a preponderance of red galaxies
nearby.

Each object in the survey is considered for its likelihood to be a BCG
in this way, using the likelihood functions to determine its likeness
to a BCG and its environment at a range of redshifts. A
maximum-likelihood redshift is assigned by
\begin{equation}
\mathcal{L}_{tot}(z_{max})=\max(\mathcal{L}_{tot}(z))
\end{equation}
(see Figure 4, vertical dotted line), and the
overall likelihood that an object is a cluster center is
\begin{equation}
\mathcal{L}_{tot}^{max}(z)=\mathcal{L}_{R}(z_{max})\mathcal{L}_{BCG}(z_{max}).
\end{equation}
The result is a list of candidate BCGs with maximum-likelihood
redshifts, $z_{max}$, and maximum cluster likelihoods
$\mathcal{L}_{tot}^{max}$. Each object on the list is marked as a
candidate cluster center, with a richness $N_{gals}$ and a scaled
richness $N_{gals}^{r200}$ (see the beginning of \S
\ref{sec:algo}). Its ``members'' list includes galaxies within a
projected separation of $r_{200}$, $2\sigma$ of the ridgeline colors, brighter
than $L_{min}$, and dimmer than the candidate BCG.

By these means, all galaxies within the specified color--magnitude
range in the survey are tested as candidate centers. In a typical rich
cluster, dozens of members will have been evaluated as potential
centers. To build the final cluster list from these candidate centers,
the candidate centers are first sorted by decreasing total maximum
likelihood. The top object on the list is taken as the BCG of a
cluster. All other objects with 1) maximum likelihoods less than this
object, 2) redshifts within $\pm 0.02$ (the approximate redshift
precision for all richnesses) and 3) within $r_{200}$ of the object,
are flagged and prevented from seeding new clusters. The next object
on the list is treated in a similar way, and by this simple
percolation process, candidate BCGs are flagged whenever they fall
within the confines of a higher likelihood object. This process
descends all the way through the list of candidate BCGs. The remaining
unflagged objects enter the final cluster catalog. In the studies
presented herein, we truncate the process at objects with
$N_{gals}^{r200}=10$ and less. While weak lensing and dynamical mass
estimates \citep{mckay06,sheldon06} indicate the lower richness
objects really do trace lower-mass systems, it is not clear what other
selection effects may be taking place below this limit.  In
particular, below the average mass for this limit, some groups may
have less well-defined red-sequences and central galaxies that have
different properties.  An important feature of the execution is that
the reported center of the cluster lands on a cluster galaxy, which
typically has the characteristic properties of a BCG.

Figure 6 shows the local value of the composite
likelihood function in a 1-degree field centered on Abell 1689. A
double peak, corresponding to two galaxies near Abell 1689's center,
is located at (ra,dec) of $(197.87,-1.34)$. The SDSS image of Abell
1689 is shown in Figure 7.  Two other
previously-identified clusters, one X-ray, one optical, fall in this
same field and are easily singled out. Their images are shown in
Figures 8 and 9. The large number of peaks with $10 <
\mathcal{L}_{tot}^{max} < 100$ correspond to group-sized objects, many
of which are absorbed as members of the higher likelihood
cluster-sized objects.

\section{maxBCG Selection Function}
\label{sec:selection}
A cluster catalog is only useful for cosmological constraints insofar
as its purity and completeness can be understood, both as a function
of redshift and of halo mass and/or halo richness. Such measurements
have typically been made in X-ray and optical cluster-finding
algorithms by Monte--Carlo methods, in which galaxies (or X-ray
photons) with various radial distributions are inserted into a
suitable background \citep{postman96,gal03}. The specific parameters
of the model are usually varied to demonstrate an insensitivity of the
measurements to the particular choice of parameters.

Here, we begin by taking the Monte--Carlo approach, following a technique similar to 
that described in \citet{goto02}. To quantify completeness, we first 
shuffle input SDSS galaxy catalogs \citep[see][]{koester06} by randomly
reassigning galaxy colors and smearing the positions by 5'. We insert
artificial clusters in the following way: 1) at five discrete redshifts, $z=0.1,0.14,0.18,0.22,$ and $0.26$, we extract photometric data for Abell clusters in the SDSS. 2) After background subtraction, we measure the average color and radial distributions of the Abell clusters stacked at these five
redshifts (\citet[][]{koester07}). 3) At each redshift, the radial distributions, are fit with power laws, and the color distributions are fit with 4th-degree polynomials. 4) To ensure a realistic richness and redshift distribution in the artificial cluster catalog, clusters are constructed by randomly choosing a redshift and richness from the actual maxBCG cluster distribution \citep{koester06}. The most nearby of the 5 discrete redshifts is chosen as a suitable model for the color and radial distributions of clusters at that redshift; k-corrections are applied to the colors so that they are actually consistent with the randomly chosen redshift (v4.1.4 of KCORRECT, \citet{bla03b}.) 5) A total of 15,000 clusters are inserted into the shuffled background, and maxBCG is run on the resulting galaxy catalogs. To determine if an artificial cluster is found by these means, we ask for each artificial cluster, ``is there a maxBCG cluster within $\pm 0.025$ in redshift and $N_{gals}^{r200} \geq 10?$ If so, the cluster is considered detected. 

In Figure 10, $90\%$ completeness is reached in all redshift ranges by $N_{gals}^{r200}=20$. The decreased completeness below this range is partially due to the fact that the artificial clusters were constructed to statistically represent rich Abell clusters, and not the poorer group-sized systems we also wish to detect with maxBCG. 

To measure the false-positive rate, we simply run maxBCG one the same shuffled galaxy catalogs constructed above. At $N_{gals}^{r200}=10$, we detect 178 clusters, compared to the 2558 in the maxBCG catalog described in \citet{koester06}, for a $7\%$ false-positive rate. At $N_{gals}^{r200}=15$, the rate is $1\%$, and is $< 1\%$ by $N_{gals}^{r200}=20$. Two clusters at $N_{gals}^{r200}=21$ are the largest systems detected in the shuffled catalogs. The false-positive rate is indeed small.

Using traditional methods to evaluate completeness and purity, the maxBCG algorithm fares very well. However, to more fully assess the catalog's quality, we employ mock
galaxy catalogs largely designed for the purpose of understanding
maxBCG selection effects, whose galaxy distribution based on the
underlying dark matter distribution and tuned to match observed
luminosity-dependent and color-dependent galaxy clustering
measurements \citep{wechsler06}. There are two motivations for
proceeding in this way. First, compared to simple Monte--Carlo
realizations, the mock catalogs are more representative of galaxy
clusters in the universe, and will reveal a more realistic picture of
the performance of the cluster finder. While any given mock catalog is
unlikely to be a perfect representation of the Universe, as we do not
understand everything about the physics of galaxy formation and galaxy
bias, the closer such catalogs can be, the more robust our
understanding of the cluster-finder selection effects can be.  These
mocks have a strong advantage over previous simplistic methods in that
they embed their clusters in a realistic network of filaments and
voids, a feature essential for understanding the effects of projection
on cluster finding. Secondly is that much of our understanding of
large-scale structure is embodied in N-body simulations, and the
strength of cosmological constraints comes from understanding the
connection of observables with dark matter halos in these
simulations.  Thus mocks that correctly encode this connection are
essential to extract the full power of these data.  Ideally, our
simulations would directly predict galaxy properties and their
distribution from first principles, but we are currently a long way
from a complete theory of galaxy formation that can do this robustly.
Empirical simulations that connect very realistic galaxy distributions
with dark matter halos are the first step towards this eventuality,
and allow us to compare observations directly to such simulations.

We establish for the remainder of this work the following definitions:
a $halo$ refers to an object in the mock catalog, consisting of dark
matter and constituent galaxies. A $cluster$ refers to any object
output by the maxBCG algorithm, whether it is run on real data or on
the mock catalog.  As part of this exercise, we use richness
definitions particular to halos and derived clusters separately. In
Table 1, we provide a summary of the various quantities we employ. The
halo-specific measurements include the mass ($M_{200}$), an intrinsic
``true'' richness of the halo, as assigned in the mock catalog
($N_{int}$), and the richness of red galaxies in these halos,
$N_{int}^{red}$. This latter quantity is aimed at describing the
richness of each halo as it would be seen by maxBCG in an ideal case:
given the halo's redshift, the number of galaxies from $N_{int}$ is
computed by creating a color--magnitude box centered on the E/S0
ridgeline colors, $2\sigma$ wide. The cluster-specific measurements,
$N_{gals}$ and $N_{gals}^{r200}$, have already been defined.

MaxBCG is run on the mock catalog in an identical way to how it is run
on the real data. We present a naive comparison in Figure 11. The halo
abundance in bins of $N_{int}^{red}$ is compared to the
$N_{gals}^{r200}$ distribution of the derived maxBCG clusters from
that simulation. For reference, $N_{int}^{red}=100$ objects have
masses of $M_{200} \sim 1\times 10^{15}$, and $N_{int}^{red}=50$
objects have masses of $M_{200} \sim 5\times 10^{14}$ In principle,
$N_{int}^{red}$ of halos is similar to $N_{gals}^{r200}$, in that they
use the same color cuts, and have a physical scale associated with
them.

The abundances are approximately consistent in their slopes, but there
appears to be an overabundance of clusters at all richnesses. This
could be due to the performance of the cluster finder itself, wherein
it tends to merge smaller systems along the same line of sight. It
could also possibly stem from small-scale projection effects that
would cause maxBCG to overestimate the halo richness, making
$N_{gals}^{r200}$ an overestimate of $N_{int}^{red}$. In the next
section, these possibilities as well as others are investigated.

\subsection{Completeness and Purity}

As part of the output of maxBCG, we include cluster members, which are
galaxies within a certain spatial distance of the BCG and inside a
color--magnitude box appropriate to the maximum likelihood redshift. As
these are predominantly red-sequence galaxies, a large fraction are
physically associated with the cluster, at around the $80\%$ level, as
indicated by spectroscopy \citep{koester06}. The members are thus
good indicators of the cluster's position in physical space, so we use
them as such when comparing to the simulations. The same goes for the
mock galaxy catalog: halo galaxy members are good indicators of the
halo positions, as they all sit with $r_{200}$ of the halo. We thus
wish to use the overlap between halo galaxies and cluster galaxies to
assert whether or not a halo in the simulation is ``matched'' to a
cluster returned by maxBCG. In the same spirit as $N_{int}^{red}$, we
only consider red halo members, i.e. the exact galaxies that make up
$N_{gals}^{red}$ are considered halo members.

In this exercise, we consider runs of maxBCG on three mock galaxy catalog realizations of the same cosmology in the interests of minimizing sensitivity to statistics. We restrict ourselves to a cluster catalog with
objects of $N_{gals}^{r200} \geq 10$, and $0.08 < z < 0.32$, as this is
the final cut applied to the catalog presented in
\citet{koester06}. The halo catalog includes halos of $M_{200} > 5
\times 10^{13}$ solar masses and redshifts $0.1 < z < 0.3$. To aid in
the interpretation, we first note that this mass limit corresponds to
an average $N^{red}_{int} \simeq 5$ and $N_{int} \simeq 10$.  The
redshift correspondence between ``matched'' (see below) clusters and
halos is quite good, but the fact that there is some scatter between
the maxBCG photometric redshift and the halo redshift (Figure 12) should be accounted for in the interpretation.

To measure the completeness, we ask the question ``for a given halo,
which cluster contains the largest fraction of the red halo
galaxies?''. Call this fraction $f_h$, and the cluster that satisfies
this criteria, the ``best'' match. We can define a threshold for $f_h$
below which we consider a halo to remain unmatched. For this exercise, we
choose $f_h$ as follows: First, note that the cluster
   membership criteria counts bright red galaxies within $r_{200}$, and in the
   simulations, halo richness is also computed inside an $r_{200}$ value
   determined from the $dark matter$ profile. When matching, only the
   bright red halo galaxies inside $r_{200}$ are considered.
   The median $r_{200}$ of the dark matter halos with
   $M_{200} > 8 \times 10^{13}$ is
   $r_{200}=0.95$ Mpc. However, that used in maxBCG for clusters with 10 or
   more red galaxies is $0.88$ Mpc, so that ratio of the areas is
   $0.88^2./0.95^2.$ or $\simeq 0.3$. Thus, $f_h=0.3$ is an appropriate threshold.

In Figure 13, we choose a threshold of $f_h=0.3$ , and
display contours of completeness across the full redshift and mass
range of halos. Evidently, maxBCG exhibits a very high completeness at
high mass, which begins to decline at around $2\times 10^{14}
h^{-1}M_{\sun}$. Many of these ``missed'' objects are assigned
$N_{gals}^{r200} < 10$, which excludes them from the richness cut we
make here, a manifestation of the mass-richness scatter. Indeed, at
$8\times 10^{13} h^{-1}M_{\sun}$, there are only about 5 red galaxies
in a typical mock halo.

It is safe to conclude that maxBCG locates
clusters with high completeness down to at least $M_{200}=2 \times
10^{14}M_{\sun}$, and certainly identifies a large population of even
lower mass objects. Thus, our stated completeness is quite
conservative; in particular if completeness were defined without this
explicit cut in richness it wouldn't decline at these masses 
(see \citealt{rozo06} for a discussion).

Figure 13 also reveals that the cluster catalog includes
halos with $z > 0.3$, nearly at the edge of the simulation. Although these halos are within the redshift error of maxBCG ($\sigma_z \simeq 0.015$,\citet{koester06}), they introduce a level of contamination to the final cluster catalog. The same could be said for lower mass halos being misidentified as higher mass halos, but the interpretation is less clear, as the mass-observable scatter is quite large; the idenification of low mass halos could still be within the limits of the mass-observable scatter without actually being contamination.

The purity is defined here by asking the reverse question: ``for a
given cluster, which halo contains the largest fraction of the cluster
members?''. This fraction is $f_c$, and the halo that satisfies this
criteria is the ``best'' halo match. A threshold at $f_c=0.3$ is
chosen for defining a match, and the purity contours in richness and
redshift space are shown in Figure 13. At redshift
edges of the catalog, the purity declines slightly. This is due to the
fact the halo catalog we match to is truncated at $z=0.3$ and $z=0.1$,
the range we are concerned with at present. Thus there are some
clusters with photometric redshifts that place them within the
redshift bounds of the cluster catalog, but whose associated halos
have true redshifts just outside this range. This is evident in the
scatter of the halo-to-cluster redshift correspondence shown in Figure 12. We can confidently say that over the range of the catalog, the
purity is well over $90\%$.

\subsection{Richness Scaling}
In the case of completeness, it is not enough to simply say that all
halos above some mass threshold were ``found''. By the matching
executed above, a massive halo with a high $N_{int}^{red}$ could be
matched to a cluster with low $N_{gals}^{200}$. While this is
considered ``found'', ideally a halo that was successfully flagged by
the cluster finder would be associated with a cluster of a richness
that is reflective of the halo's richness. In this sense, we can take
the matching exercise a step further and ask ``how well does the
richness of the best matched cluster reflect that of the underlying
halo?''

In the top panel of Figure 14, halo richnesses and the
richness of the best-matched cluster are compared for one of the mock galaxy catalogs. Each halo is
represented by a cross or a diamond. Halos represented by crosses are
those that were the most massive match to each cluster. Halos
represented by diamonds are those that were matched with clusters that
had already been previously matched by a larger halo. Following
\citet{gerke05}, we call this ``overmerging''. There is a large
population of halos with well-matched clusters, i.e., clusters with
richnesses indicative of the underlying halo properties. However,
there is a subset of overmerged halos (diamonds) with low
$N_{int}^{red}$ and high $N_{gals}^{r200}$. Naively, this may be
attributed to catastrophic projection effects that have been known to
plague optical cluster finding. This in turn could mean that we have
clusters we think are rich, but really are just projections of many
random things along a deep line of sight. But there is another
possibility. Our likelihoods will not resolve halos within a few Mpc
of each other, along the line of sight, particularly large halos with
a few smaller neighbors. Thus maxBCG can merge these objects,
selecting one large, dominant halo and merging the smaller
neighbors. Members from these neighboring halos overlap strongly with
the cluster members from the dominant cluster, so all the halos match
to the $same$ cluster. This explains the low $N_{int}^{red}$, high
$N_{gals}^{r200}$ population seen in Figure 14.

The three boxed points at the of the top panel of Figure 14 illustrate this merging. They are three separate
halos of masses $1.03 \times 10^{15},$ $6.57 \times 10^{14},$ $3.94
\times 10^{14}M_{\sun}$, $N_{int}^{red}$ of 89, 62, and 38, redshifts
at 0.233, 0.229, and 0.235. The latter two appear as diamonds. These
halos are in fact all within $ < 7.1h^{-1}$ Mpc of each other,
approximately along the same line-of-sight. This physical distance is
below the resolution we expect to gain from the red sequence. The
cluster that is best matched to all three of these halos has
$N_{gals}^{r200}=269$, greater than the sum (189) of these three
halos. A large portion of the objects in this region of Figure 14 are less severe instances of this same type of
overmerging. But the matching algorithm used here generally associates
rich clusters with rich halos, so our quoted completeness is dually
informative. These smaller halos (diamonds) are evidence that maxBCG
is performing as expected, merging smaller, nearby halos with larger
ones.

To further elucidate the situation, the question can be turned around:
``how well does the richness of the best matched halo reflect that of
the underlying cluster?''. This asks the same question of
cluster-to-halo matching for the purity measurements. In the lower
panel of Figure 14, for each cluster, we plot cluster
$N_{gals}^{r200}$ and the $N_{int}^{red}$ of the best matched
halo. This selects one best halo for each cluster. The excess at low
$N_{int}^{red}$ and high $N_{gals}^{r200}$ seen in the top panel
vanishes, which shows that indeed those small halos(diamonds in the
upper panel) were not poorly matched, but that they were absorbed into
large clusters. The case described above of extreme overmerging shows
up as one point at (89,269).

The diamonds that appear on the lower panel are clusters that were
matched to a halo which had already been claimed as best match by a
richer cluster. This $fragmentation$, in which the halo is broken up
by maxBCG, is much less prevalent than overmerging (this is quantified
in the following section). When fragmentation does occur, there is
still generally one large halo matched to one large cluster, with a
few smaller clusters in the neighborhood as well.

The most interesting case to study here is the two clusters with
$N_{gals}^{r200}=223,15$, and redshifts $0.170,0.173$. These two
correspond to one halo, with a mass $M_{200}=1.726 \times
10^{15}M_{\sun},N_{gals}^{red}=163$, and $z=0.167$. There is one clear
dominant cluster here, and two others with much smaller
richnesses. This is a case of fragmentation, but it is not especially
problematic because there is still a dominant cluster associated with
the halo. This same dominant cluster shows up in the top panel of
Figure 14 at (163,223).

It is safe to conclude then that we are successfully matching rich
clusters and rich halos, and that our completeness and purity
measurements are telling us much more than ``did we find all the
halos'' and ``is there a halo there or not?'' By either means, rich
halos are associated with rich clusters. We note that this is
encouraging for mass calibration. The extension of these results to
understanding the mass-richness relation is enticing, and is
undertaken using observational weak lensing and dynamical measurements
in other works \citep{mckay06,sheldon06,koester06}.

We conclude this section by applying the insight gained from looking
at the richnesses to the abundances seen in Figure 11. In Figure
14, we display best-fit lines to richness
relations. If the offset in the abundances is really due to a
disagreement between the maxBCG $N_{gals}^{r200}$ richness and the
intrinsic $N^{red}_{int}$, and the line is a good fit, then the line
should should supply the transformation between the two, using either
fit. In the upper panel, the the best-fit line has a slope $1.409 \pm
0.013$ and intercept $1.227 \pm 0.190$. This tells us how we should
transform $N_{int}^{red}$ for the halo abundance to
$N_{gals}^{r200}$. If it is perfect, it should shift the halo
abundance curve onto the maxBCG abundance. In the lower panel, the
best-fit line has slope is $0.601 \pm 0.005$ and the intercept $1.028
\pm 0.160$, telling us how to transform the $N_{gals}^{r200}$ in the
maxBCG abundance to $N_{int}^{red}$. If this transformation is
perfect, it should shift the maxBCG abundance into agreement with the
halo abundance. Overplotted on Figure 11 are resulting abundances
under the transformation of $N_{int}^{red}$ to $N_{gals}^{r200}$
(upper dotted line and upper panel of Figure 14) and
the reverse transformation (lower dotted line and lower panel of
Figure 14). The latter transformation nearly brings
the maxBCG abundance into agreement with the intrinsic halo abundance,
while the former transformation brings the halo abundance into
agreement with maxBCG only at low richness. From Figure 14, the duplication present in the upper panel alerts
us to the difficulty in seeking a simple relation between the
richnesses, but the relation is more straightforward from the reverse
situation in the bottom panel.  Without further modeling of this
relationship, we can assert now with confidence that the offset in the
abundances observed in Figure 11 is largely due to differences in the
richness measurement.

\subsection{Fragmentation and Overmerging}
Measurements of the cluster abundance function or of cluster
clustering are potentially powerful cosmological tools. A standard
assumption is to assume that each cluster from the cluster catalog can
be matched uniquely to one halo. The extent to which this is true
depends on the cluster finder, and the definition of the halo
\citep{kim02,gerke05}. For instance if the cluster finder assigns two
clusters to each halo, an inflated abundance function will result, as
well as a correlation function with increased amplitude. A full
treatment of this relationship depends on the exact scientific goal
one wishes to address. We explore this question in detail, as it
applies to cosmological constraints from the cluster abundance
function in \citet{rozo06}. In this section, some relevant issues for
such modeling are described.

Up to this point, we have taken the ``best'' match of halos to
clusters and vice-versa to measure the purity and completeness
respectively. Using these methods, it is possible for a halo to have,
say, $70\%$ of its members come from one cluster, $20\%$ from another,
and the remaining $10\%$ not associated with any cluster. In cases
where the fractions become nearly equal, say, $50/45/5$, there is no
clear one-to-one association. This was touched upon in 14, where our matching prescription revealed examples
of halos being merged into several clusters, and a few of the
opposite, where halos were broken up into two or three clusters.

To evaluate the fragmentation incidence, we take a halo and look at
the top two clusters that contributed members, the ``best'' and the
second best. We create a simple fragmentation diagnostic:
\begin{equation}
\phi_h=\frac{N^h_2}{N^h_1}
\label{eq:frag}
\end{equation}
where $N^h_1$ and $N^h_2$ are the numbers of members contributed by
the best and second best cluster matches. $\phi_h=0$ when $N^h_2=0$
and there is no fragmentation, and $\phi_h=1$ when the halo was
perfectly broken in two. This distribution is shown as a function of
$N_{int}^{red}$ in the top panel of Figure 15. To understand the
distribution better, we bin the halos by richness and report the
median of $\phi_h$ in each richness bin (solid line) It quickly
approaches 0, indicating that there is a minimal amount of
fragmentation. The fragmented halo at $N_{gals}^{red}=163$ shows up as
a point at (163,0.2) in this panel. Out of 112 halos with
$N_{gals}^{red} > 50$, 7 have $\phi_h > 0.1$, or 6\%. For
$N_{int}^{red} > 10$, this rises to 25\%.

The exercise is repeated for clusters, considering the top two halo
matches. The overmerging statistic, $\phi_c$ is shown in Figure 15, lower panel. The median hovers around 0.2, indicating that there is some over-merging
present in the catalog at all richnesses. It is clear
that there is a population of clusters that consist of overmerged
halos. For $N_{gals}^{r200}>50$, $118/299$ halos have $\phi_c > 0.2$,
or 39\%. The number is slightly lower for the full range of
$N_{gals}^{r200}>10$, at 37\%. It thus remains true that imaging
based-cluster surveys will be subject to overmerging, as colors have
difficulty resolving of order 10 Mpc line-of-sight distances. This is another
contribution to the apparent overabundance seen in Figure 11.

More generally, rich clusters with large apertures will be particulary subject to overmerging of systems along the line-of-sight. This is likely true of any imaging survey, and future cosmological analyses using clusters from imaging surveys will require a detailed understanding of the extent to which this influences the final constraints.

\subsection{Improved modeling of selection effects}
Thus far, we have presented the basic features of the algorithm one
must incorporate into any cosmological analysis. The matching routines
executed herein are designed simply, so that we can understand the
output of maxBCG. However, consideration of the richness estimates and
the prevalence of fragmentation and overmerging using the maxBCG
algorithm demonstrates that these effects must be modeled in detail to
fully understand the selection function and properly model abundances.
\citet{rozo06} employ a matching algorithm similar in spirit to that
presented here, that makes cluster-to-halo associations clearer, and
makes the selection function of maxBCG easier to quantify. Figure 15 demonstrates the improvement one can realize in
the matching. We refer to this as ``exclusive'' matching, and point the reader to \citep{rozo06} for a full description. Note that exclusive matching does substantially better
than the ``one-way'' matching used in the current work, and in particular that the slope between the
cluster abundance and the halo abundance is now roughly unity.  We
refer the reader to this paper for a detailed discussion of how to
calibrate the full connection between clusters and halos in a way
approprate for modeling cluster abundances.

\section{Discussion}
We have described herein the key components of the maxBCG cluster
finder and their origins. The performance of its likelihood functions
have been demonstrated in a few case studies of observed galaxy
clusters. Over the redshift range of $0.1 < z < 0.3$, its selection
function was evaluated as a function of halo mass and cluster
richness. The results of the selection function tests were further
supported by the demonstration that the maxBCG cluster--mock halo
correspondence behaves as expected: rich clusters typically correspond
to rich, massive halos. These same comparisons proved to be efficient
diagnostics of fragmentation and overmerging. It seems that in
addition to the $N_{gals}^{r200}$ richness measurement being an
overestimate of richness (Figures 11 and 14), maxBCG tends to
overmerge systems along the line of sight (Figures 14 and 15).

\subsection{Features of MaxBCG}
This algorithm differs from other photometric cluster finders in its
use of likelihoods. Earlier methods have used a range of methods to
compute cluster likelihoods and assign significances
\citep{postman96,gladders00,gal00}. In these methods, cluster
candidates have been selected by assessing the global significance of
their likelihoods and then choosing some likelihood threshold above
which cluster candidates are selected. These highest significance
clusters are then reported in the catalog.

In maxBCG, we set out to push farther down the abundance function by
making no such cut. The likelihood function is used locally to report
the highest likelihood objects in some region of sky, and then the
likelihoods are used to excise locally less significant objects. At no
point do we enforce a global likelihood cut, so that the likelihood
function is permitted to define objects down to very low
richness. Although the completeness and purity are decreased in this
regime, the average properties of these lower likelihood, lower
richness objects are very encouraging. Their stacked velocity profiles
\citep{koester06,mckay06} and weak lensing measurements
\citep{sheldon06} indicate that they are indeed massive objects whose
richnesses are strongly coupled to the underlying mass
distribution. This opens up this possibility of leveraging the larger
statistics of this lower mass population in constraining cosmology, if
the selection effects can be properly understood.

Another fundamental difference in the execution of this algorithm is
its reliance on galaxy positions. We evaluate the likelihood function
at the locations of galaxies, and not on pixelized data. In
pixelization schemes, the data is broken up spatially, and likelihood
functions are tested at each pixel. Various means are then used to
assess significant peaks in the likelihood function across pixels in
the survey, and significant pixels are then tied to clusters. This
method helps to speed up the cluster search, and it also prevents
over-identification of substructure, or multiple identifications of
the same cluster. At the cost of computational time, we find that
using galaxy positions allows us to report cluster positions centered
on visually-identified BCGs. By our percolation scheme, we do not
suffer from over-identification of substructure, which would manifest
itself in our fragmentation measurements.

\subsection{Background}
Recall that we measure cluster properties down to $0.4L_* \equiv L_{min}$ at each redshift. Ideally, the $L_{min}$ cutoffs could be derived empirically from the
LRG distribution in the way that the colors were extracted. This
requires a full treatment of the luminosity function of LRGs
vs. redshift. Obviously, biases in the model could bias the richness
measurements and possibly the likelihood functions. The two major
sources of bias are incorrect $k$-corrections and, to a lesser extent,
evolution of the luminosity function.

We can take a quick look at the consequences of biases introduced by
slight errors in the offset $a(z)$ of the $a(z)+M_*$ cutoff (see
section 2.5). First, at $z=0$, we can integrate down the cluster
luminosity function \citep{hansen05} to $a(z)+M_*=-20.25$ in the
cluster's rest frame. This gives us a count of the number of galaxies,
$N_{c}$ in the cluster brighter than -20.25. Now, imagine we place the
same cluster at $z=0.2$, where the $i$-band samples a bluer, and
possibly dimmer part of the rest frame spectra of the cluster
galaxies. $K$-corrections use a model of the spectra of such galaxies to
correct for this systematic offset, but the model may represent this
offset with varying degrees of success at different redshifts. Assume
that the $k$-corrections at $z=0.2$ are too dim by only $0.15$ magnitudes
on average. Call this error $\delta m$. We can again integrate down
the same luminosity function, this time to $0.15+a(z)+M_*$, and get a
number of cluster galaxies $N_{c}+\delta N_c$. Comparing this value to
that of the rest-frame ($N_{c}$) reveals that the number counts can be
$20\%$ higher when the $k$-corrected limiting magnitude $L_{min}$ is
$0.15$ too dim. The fractional errors in the number of cluster
galaxies for a given magnitude error, $\delta m$, can be cast in the
following form:
\begin{equation}
\frac{\delta N_c}{N_c} \simeq 1.33\delta m
\end{equation}
Thus, when $\delta m$ varies with redshift, it is clear that the
richness estimate and the likelihood function (see next paragraph) can
have unwanted redshift dependence built in.

If there are indeed biases in $L_{min}$, it will certainly affect our
richness measurements, but we don't expect it to bias the cluster
finder itself.  First, because the $\mathcal{L}_{R}$ varies slowly
with redshift the backgrounds at adjacent redshifts, say $z=0.2$ and
$z=0.21$, are very similar. Second, the redshift-color relation is
smooth and well-understood so that the color filter component of the
likelihood function does not radically change. Last, the ridgeline
likelihood function is usually strongly peaked as a function of
redshift, so that there is an obvious maximum in redshift
space. Again, this is not a problem for rich objects, because the
detections are robust. Redshift biases are more likely to be seen in
low richness objects, where the detections are not high S/N and the
background number counts are not well-approximated as Gaussian, and
the galaxy population is not as dominated by red-sequence
galaxies. Broader spectroscopic samples and realistic mock galaxy
catalogs that include more faint galaxies will enable a more
systematic study of the effect of background in maxBCG cluster
detection.

If there are obvious redshift biases introduced by this method, in
particular the model for $L_{min}$, they will be borne out in the
catalog (see Paper I). In particular, for $0.1 < z < 0.3$, the number
of objects should increase like the volume enclosed. In the
accompanying paper \citep{koester06} we demonstrate that this is
approximately the case, and the catalog is volume-limited to z =0.3.

\subsection{Model Biases}
There are notable exceptions to our model for the BCG colors. A
well-known population of so-called cooling-flow clusters contain BCGs
where gas is in the process of cooling and forming stars, which
creates a significantly bluer BCG. The presence of emission lines in
the BCG spectrum provides evidence for a cooling flow. A1835 is a
well-known example \citep[e.g.][]{mcnamara06}. While we successfully find such clusters,
we do not pick the brightest cluster galaxy as the center because its
$g-r\simeq 0.6$ is significantly bluer than the other red-sequence
cluster galaxies, which all hover around $g-r\simeq 1.3$. In the BCS
X-ray cluster catalog \citep{ebeling98}, $27\%$ of BCGs show some sort
of emission lines (Crawford et al., 2003). The extent to which
emissions lines are prevalent in BCGs living in the cluster population
as a whole, and the impact they have on the BCG colors, are questions
we can begin to address with maxBCG-selected catalogs. The study of
\citet{weinmann06}, which selects low-redshift groups from
spectroscopy without reference to color, hints that BCG colors
different from those of the cluster population are the exception
rather than the norm.

The matched-filter technique applied here is built around the colors
of the E/S0 ridgeline and the NFW distribution expected in clusters of
galaxies. The idea is that it should faithfully describe the average
properties of galaxy clusters. Certainly, it will discriminate against
objects that don't fit this model, which is more likely to be true at
lower mass and richness, and potentially at higher redshift.  We can
only quantify this in so far as the mock catalogs are a realistic
representation of the cluster and background populations.  However, it
is worth pointing out that any cluster finder that operates on photometric
data will have to make some assumptions about the galaxy populations
and their distributions, and it is thus worth making these assumptions as easy to understand and as close to reality as possible.

A final note is that the cluster finder is requires an assumption
about the cosmological dependence of the angular diameter distance and
the cluster luminosity function used at various stages of the
algorithm. Given a large enough sample with redshifts, the luminosity
function could be determined observationally, but for large changes in
cosmology, it is less clear what to do about the angular diameter
distance. The sensitivity of both of these to cosmology is currently
under investigation.

\subsection{Photometric Errors}
As the redshift of an object increases, so does its photometric
error. One would like for the photometric errors to be much less than
the intrinsic ridgeline width. As the errors get bigger, we allow
galaxies to be tested as members in more clusters. This can
potentially affect the richness by preferentially inflating richnesses
at high redshift, where the color errors are larger. The cluster
catalog presented in \citet{koester06} is approximately volume-limited
in all richness bins, which suggests that this effect is not too
severe. However, a more well-understood richness measurement comes at
the price of the information lost by discarding galaxies with large
errors. Appropriately dealing with photometric errors is currently an
issue of interest.

\subsection{Issues for extending to higher redshift}
The SDSS is a goldmine of cluster data, and it provides an important
testing ground for algorithms which will generate the cluster catalogs
important to cosmological constraints in future surveys. We have
restricted ourselves to a redshift range where the clusters are
well-measured, and most of the redshift and color information is
contained in one band.

The selection is quite uniform across all redshifts, but this may
change when the search encompasses $z > 0.35$, as the 4000 \AA\ break
migrates in the $r$-band, and $r-i$ contains more information. Future
surveys will aim to find clusters uniformly across a broad range of
redshifts, spanning many filters. The likelihoods that embody these
color models must weight the different colors fairly, and smoothly
handle transitions from one band to another.

The SDSS is deep enough that the same methods provide a means
to select clusters out to at least $z \sim 0.6$, and still see the
brightest members. Toward this redshift, the flux-limit of the survey
becomes an issue in fairly measuring richnesses and evaluating
likelihood at all redshifts.

\subsection{MaxBCG vs. Other Wavelengths}
The results presented here and in \citet{koester06} indicate that maxBCG recovers galaxy clusters with purity and completeness levels of $90\%$ for $M > 2 \times 10^{14} M_{\sun}$ and $N_{gals}^{r200} \geq 10$. X--ray selected clusters of similar sky and redshift coverage such as the Brightest Cluster Sample \citep[BCS][]{ebeling98} and the Northern ROSAT All-Sky \citep[NORAS][]{bohringer00} Survey are complete at the $80-90\%$ level above a given flux limit which will naturally impose a selection effect on the resulting cluster sample that is strongly dependent on redshift. Redshift-dependent effects in maxBCG arise only through biases in $L_{min}$ over $0.1 < z <0.3$. \citet{bohringer00} reach $76\%$ in purity, as determined through optical follow-up of X--ray selected clusters. The purity of X--ray samples is altered by unresolved active-galactic nuclei or stellar contamination, while the purity of optical samples can be compromised by projection.

Aside from the completeness and purity of the sample, understanding the mass distribution of the cluster sample is of utmost importance. From the preceding analysis, it is clear that projection plays a role in both richness estimation and in overmerging. These can both bias the estimate of the mean mass at fixed richness, and the estimate of the scatter. The same can be said of mass estimates in X--ray and SZ surveys: point-source contamination, mergers, and other gasdynamical processes contribute to the overall X--ray signal. No technique is perfect, but for given cluster sample, the extent to which contamination of the observables can be understood will strongly affect the resulting cosomological parameter constraints.

\section{Summary}
The maxBCG algorithm represents a significant step forward in large
galaxy survey cluster detection. Much is known about the properties of
cluster galaxies, and the availability of immense, rich, imaging
catalogs allows us to begin leveraging this knowledge to detect
cleaner and more robust cluster samples and to provide new
cosmological information. Since optical surveys in the near future
intend to pursue cluster cosmology to varying degrees, the advances
presented here are timely. We summarize them as follows:
\begin{itemize}
  \item{MaxBCG pushes beyond the high end of the abundance function,
    down to group-sized halos, where there is additional cosmological
    information. The purity and completeness across this broad range
    are above $90\%$ for $N_{gals}^{r200} > 10$ and $M_{200}>2 \times
    10^{14}$, respectively, across $0.1<z<0.3$. Systems below this
    mass range are quite accessible as well, depending on where one
    draws the richness cut.}
  \item{In addition to spatial clustering of red-sequence galaxies, we
    add in information about brightest cluster galaxies and multiple
    colors to refine the search.}
  \item{The likelihoods are evaluated at individual galaxies, as
    opposed to pixelizing the sky.}
  \item{The selection function of the algorithm is explored in detail
    with mock galaxy catalogs. Since the mock catalogs contain a
    wealth of observationally-motivated information that is coupled to
    the underlying mass, we are able to undertake an unprecedented study
    of the performance of the algorithm on individual halos to gain a
    deeper perspective on the halo selection. }
  \item{The mock catalog provides an initial demonstration of the
    richness--mass mapping; richer systems are preferentially
    associated with more massive halos.}
  \item{We confirm the expectation that on $\sim 10h^{-1}$ Mpc scales,
    maxBCG tends to overmerge systems projected along the
    line-of-sight, an effect that will have to be
    modeled. Fragmentation is not found to be significant.}
  \item{We open the discussion of difficulties that will be
    encountered in future multi-band, wide-angle, high-redshift
    cluster surveys. Among these are uniformity of the richness
    measurements, challenges in using multiple colors, photometric
    errors, and quantification of the selection function.}
\end{itemize}

In an accompanying paper \citep{koester06} and others soon to follow,
maxBCG is further assessed with $\simeq 7500$ deg$^2$ of SDSS imaging
data. In this test on real data, the scaling of cluster richness with
velocity dispersion is demonstrated, as are the quality of photometric
redshifts, the incidence of projection, the uniformity of richness
measurements and the agreement between these optical clusters and
earlier X-ray selected catalogs. The maxBCG method provides a
different approach to optical cluster selection with a number of
useful features. The experience gained in using it on SDSS data will
provide important guidance in planning and executing future surveys,
such as those planned to study dark energy, including the Dark Energy
Survey \citep{des05}, the Panoramic Survey Telescope \& Rapid Response
System \citep{kaiser05}, the Supernovae/Acceleration Probe
\citep{snap02}, and the Large Synoptic Survey Telescope
\citep{haiman04}.

\acknowledgments
Funding for the SDSS and SDSS-II has been provided by the Alfred
P. Sloan Foundation, the Participating Institutions, the National
Science Foundation, the U.S. Department of Energy, the National
Aeronautics and Space Administration, the Japanese Monbukagakusho, the
Max Planck Society, and the Higher Education Funding Council for
England. The SDSS Web Site is http://www.sdss.org/. T. McKay,
A. Evrard, and B. Koester gratefully acknowledge support from NSF
grant AST 044327. R. Wechsler is supported by NASA through Hubble
Fellowship grant HST-HF-01168.01-A awarded by the Space Telescope
Science Institute.  We are grateful for the repeated hospitality of
the Aspen Center for Physics and the Michigan Center for Theoretical
Physics.

%\bibliography{bkoester}

\clearpage

\begin{deluxetable}{cll}
\tablecolumns{3}
\tablewidth{0pc}
\tablecaption{Richness Definitions}
\tablehead{
\\
\colhead{Name} & \colhead{Description} }%
\startdata
$M_{200}$ & Underlying halo mass in mock catalog\\
$N_{int}$ & Occupation number of halo\\
$N_{int}^{red}$ & Number of red $N_{int}$ galaxies, brighter than $L_{min}$\\
$N_{gals}$ & Number of red galaxies in a cluster, inside $h^{-1}$ Mpc\\
$N_{gals}^{r200}$ & Number of red galaxies in a cluster, inside $r_{200}$\\
\enddata
\tablecomments{This table is a compilation of the richness measurements used in the halo to cluster comparison in Section 4.}
\label{mmatchtab}
\end{deluxetable}

\clearpage

%\begin{figure}
%\begin{center}
%\rotatebox{90}{\scalebox{0.7}{\plotone{f1.eps}}}
%\figcaption{NFW
%  likelihood profile at different scale radii. From top to bottom,
%  the likelihood profile is shown for $r_s =$ 250, 200, 150, 100, and 50 kpc.}
%\label{fig:nfw}
%\end{center}
%\end{figure}

\begin{figure}
\begin{center}
\rotatebox{0}{\scalebox{0.7}{\plotone{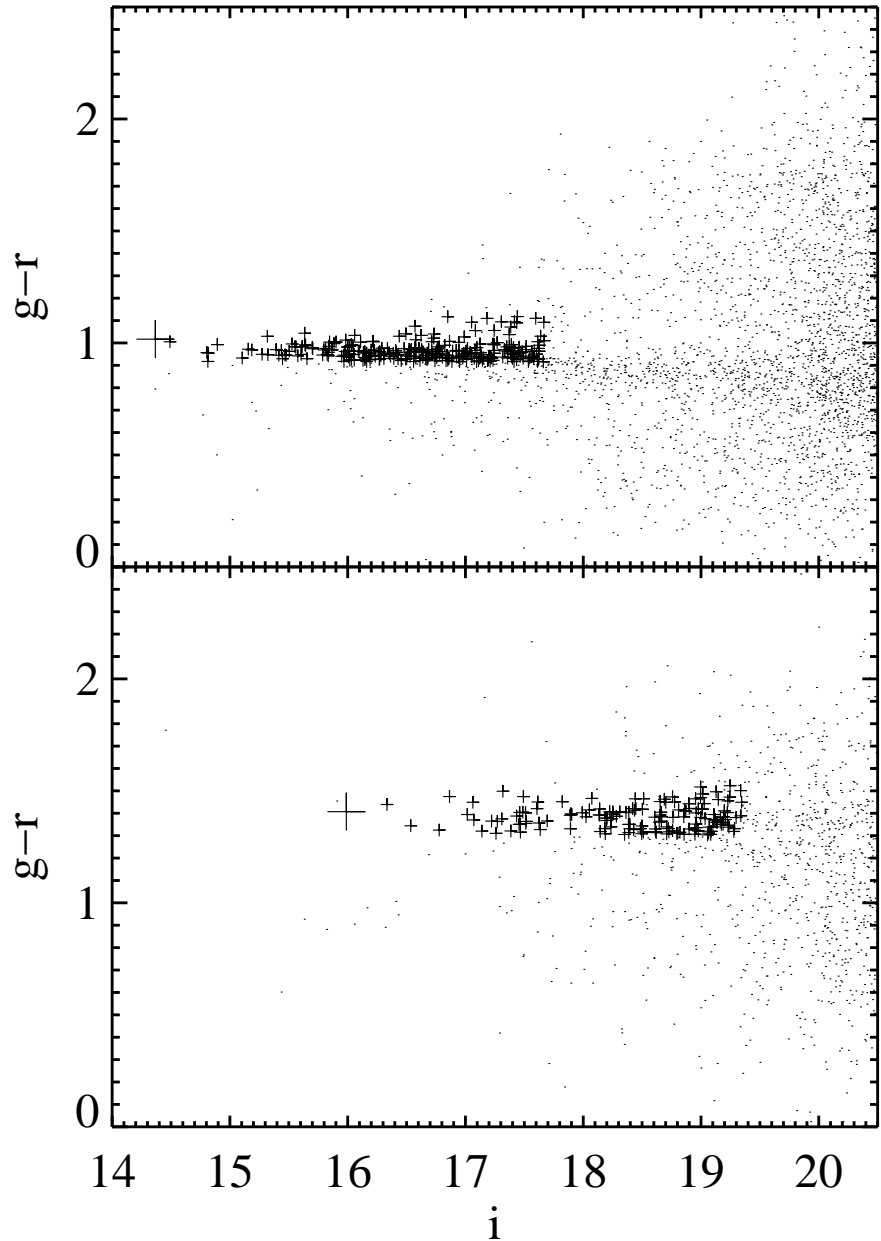}}}
\figcaption{BCGs and cluster members in the fields of two rich clusters. The
  top and bottom plots are the CMDs in the $2h^{-1}$ Mpc surrounding
  fields of Abell 2142 ($z=0.092$), and Abell 1682 ($z=0.23$). The
  large cross is the BCG, small dots are field galaxies within
  $2h^{-1}$ Mpc, small crosses are cluster members (see text). The
  small scatter and tilt in these are clear, particularly in Abell
  2142, where the ridgeline can be observed to much fainter
  magnitudes. Already at $z=0.23$, we see that the ridgeline is
  broader, primarily due to photometric errors.}
\label{fig:ridgeline}
\end{center}
\end{figure}

\begin{figure}
\rotatebox{90}{\scalebox{0.7}{\plotone{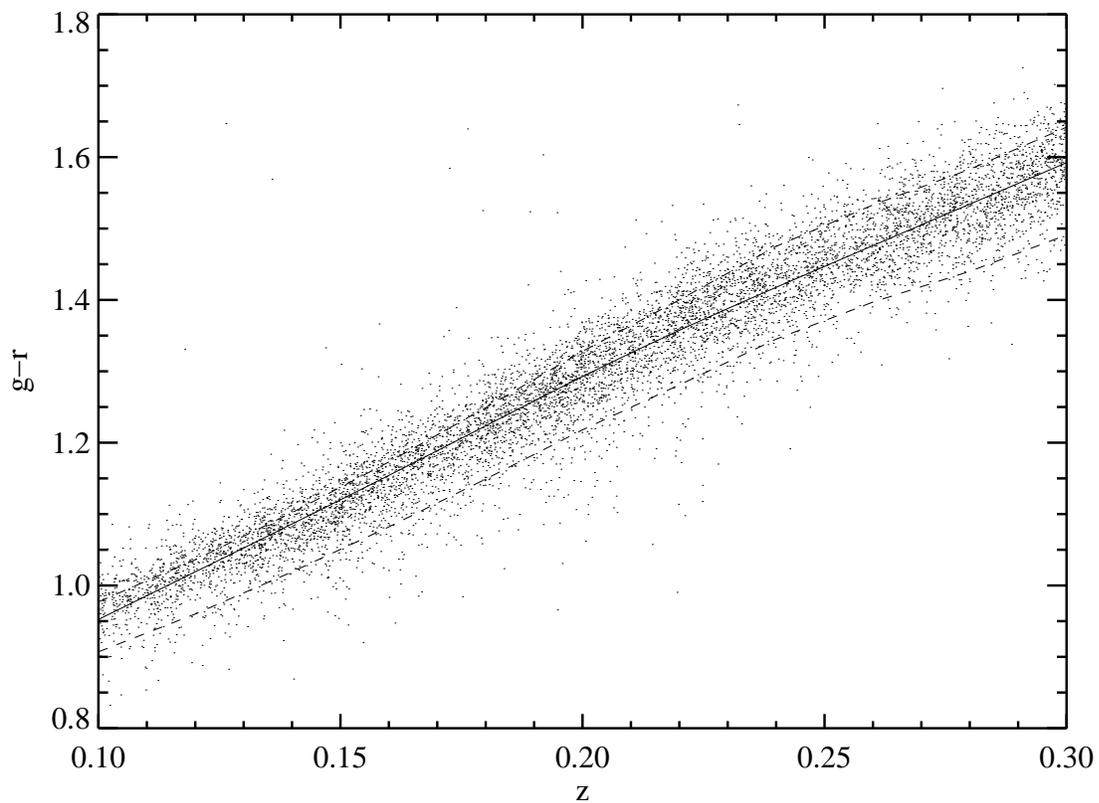}}}
\figcaption{Color ($g-r$) vs. spectroscopic $z$ input relation for clusters of
  all richnesses. Each dot represents cluster members with
  spectroscopic redshifts. Their color SDSS ($\tt{MODEL}$ magnitudes) 
  and redshift are plotted. The
  curve is the piecewise-defined ridgeline-redshift relation,
  essentially best fits to the cluster colors and redshifts. The upper and lower
  dotted lines denote the LRG passively-evolving and star-forming colors (see text).}
\label{fig:colorz}
\end{figure}

\begin{figure}
\begin{center}
\rotatebox{90}{\scalebox{0.7}{\plotone{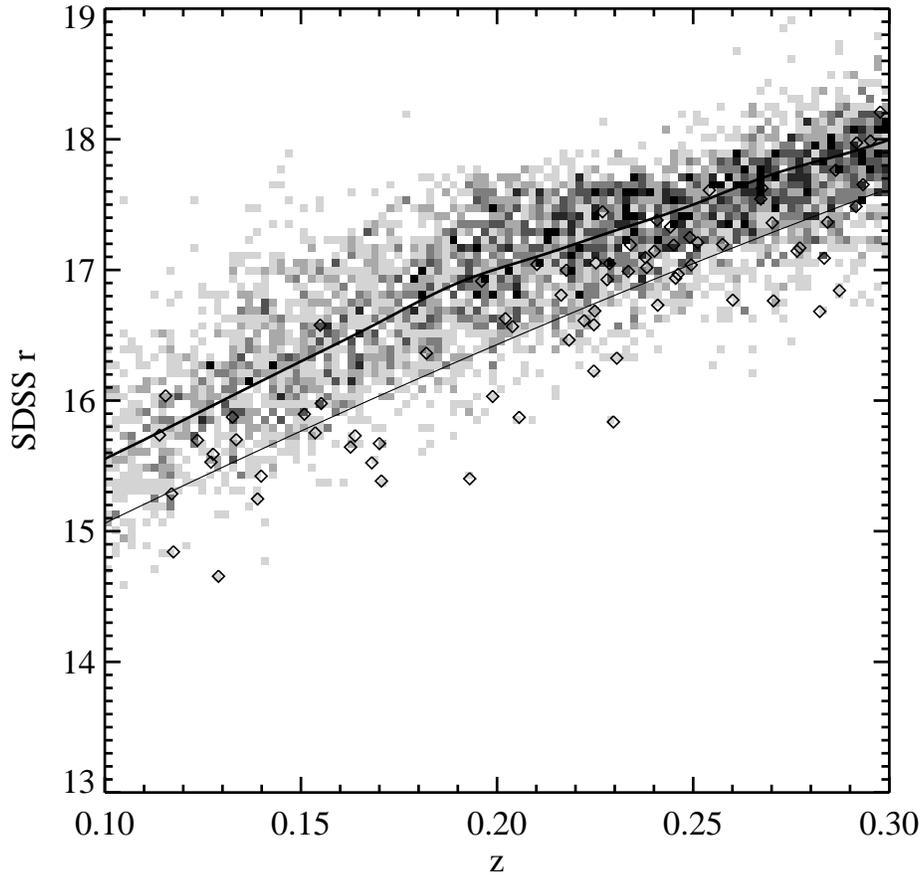}}}
\figcaption{Magnitude--redshift relation for BCGs. The upper solid curve
  is adapted from \citet{loh06}, using $r_{petro}$. The lower curve is
  the fit to the $r$-bands of 100 visually identified BCGs from a first
  run of maxBCG, diamonds on the plot. Because the magnitudes are
  observed, distance modulus, $k$-corrections, and evolution are
  automatically included. The difference between \citet{loh06} and
  this curve is due mainly to the fact that the former do not restrict
  themselves to the brightest BCGs. The greyscale density plot is the
  population of BCGs from the maxBCG catalog described in
  \citet{koester06}}
\label{fig:magz}
\end{center}
\end{figure}

\begin{figure}
\rotatebox{90}{\scalebox{0.7}{\plotone{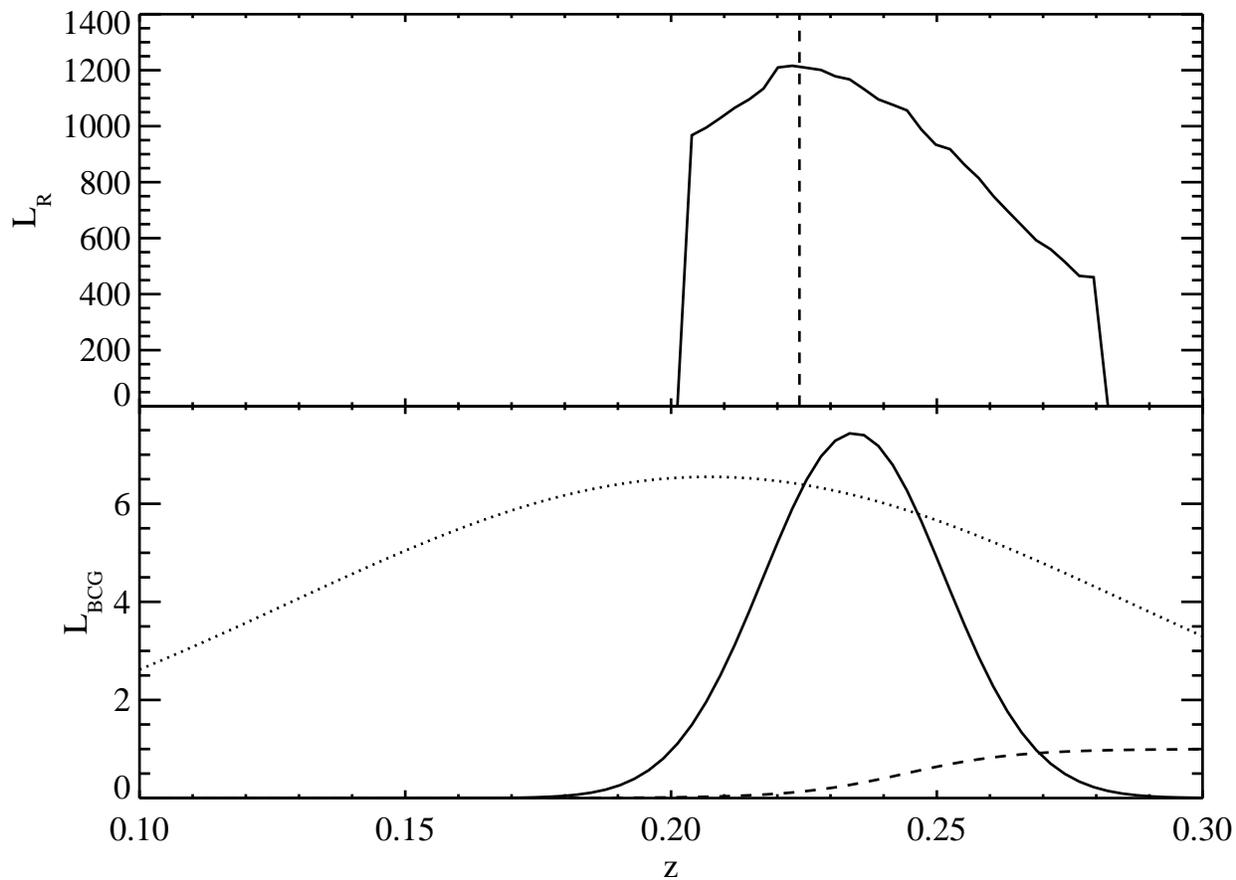}}}
\figcaption{Likelihoods for a previously unknown $N_{gals}=30$, $z=0.23$ maxBCG
  cluster. The upper panel shows a sample $\mathcal{L}_{R}(z)$, with
  the maximum-likelihood redshift denoted with a vertical line. The
  components of $\mathcal{L}_{BCG}(z)$ are shown in the lower
  panel. The narrow Gaussian is $G_{gr}$, the broad is $G_{ri}$, and
  the dashed curve is the magnitude threshold. See the associated
  image (Figure \ref{fig:image1})}
\label{fig:likelihoods}
\end{figure}

\begin{figure}
\begin{center}
\rotatebox{0}{\scalebox{0.8}{\plotone{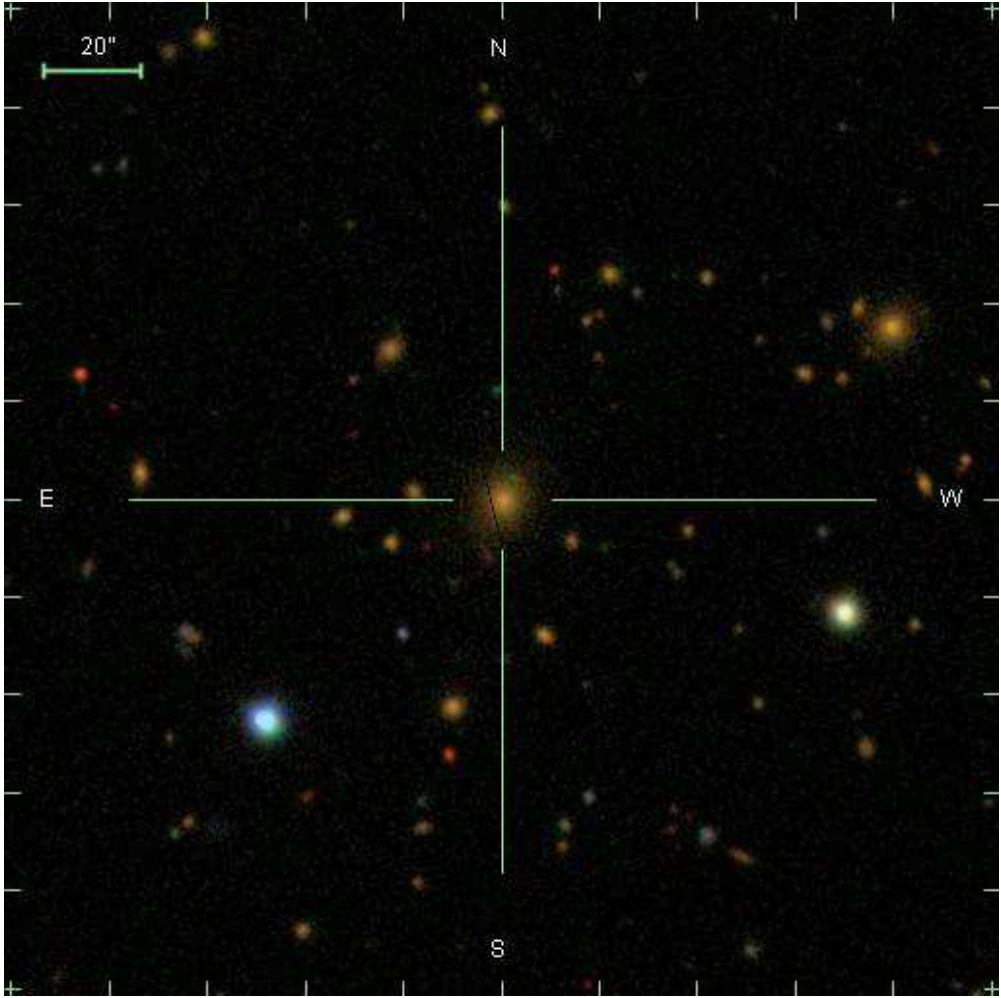}}}
\figcaption{Image of object described in text and in Figure 4. Uniform red galaxies dominate the image,
  which spans 200'', or $\simeq 0.5h^{-1}$ Mpc. See online edition of the Journal for color image. }
\label{fig:image1}
\end{center}
\end{figure}

\begin{figure}
\rotatebox{90}{\scalebox{0.6}{\plotone{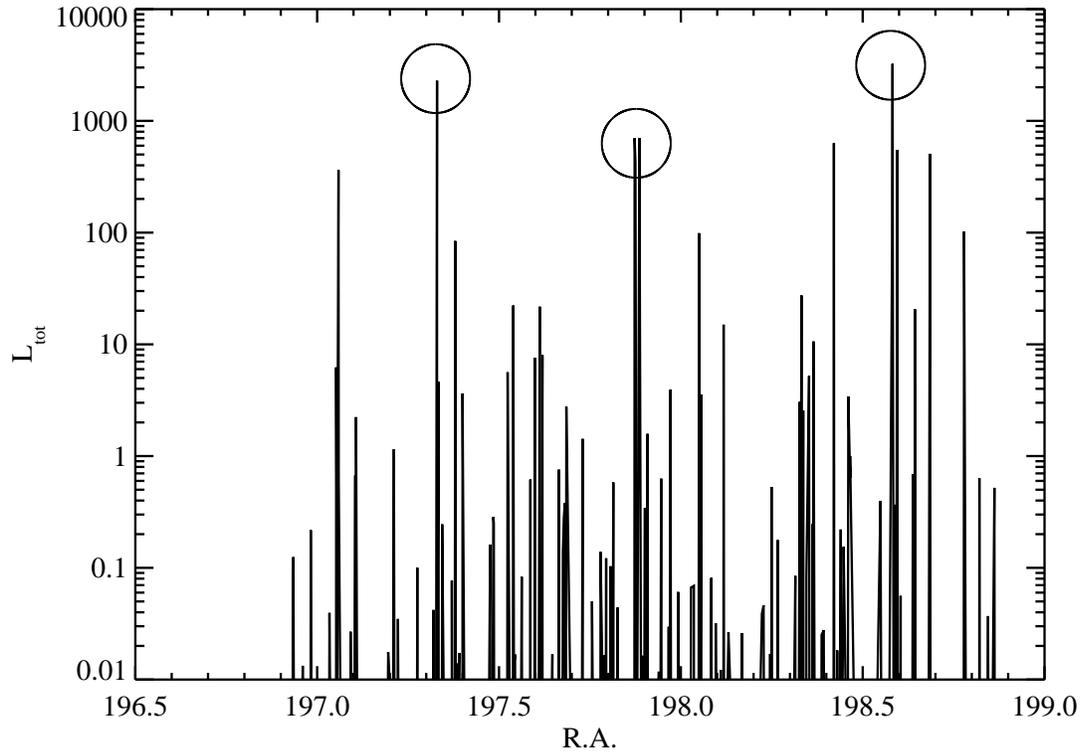}}}
\figcaption{Cluster likelihoods in the field of Abell 1689 (center
  circle). Two other high likelihood peaks also reside in this
  field. At (197.33,-1.62), $z=0.08$, is a REFLEX X-ray selected cluster,
  MS 1306.7-0121 (left circle, \citealt{bohringer04}). At
  (198.58,-1.46), $z=0.18$, is NSC J131423-012734, part of the Northern
  Sky Optical Cluster Survey \citep{gal03}.}
\label{fig:abell_like}
\end{figure}

\begin{figure}
\begin{center}
\rotatebox{0}{\scalebox{0.8}{\plotone{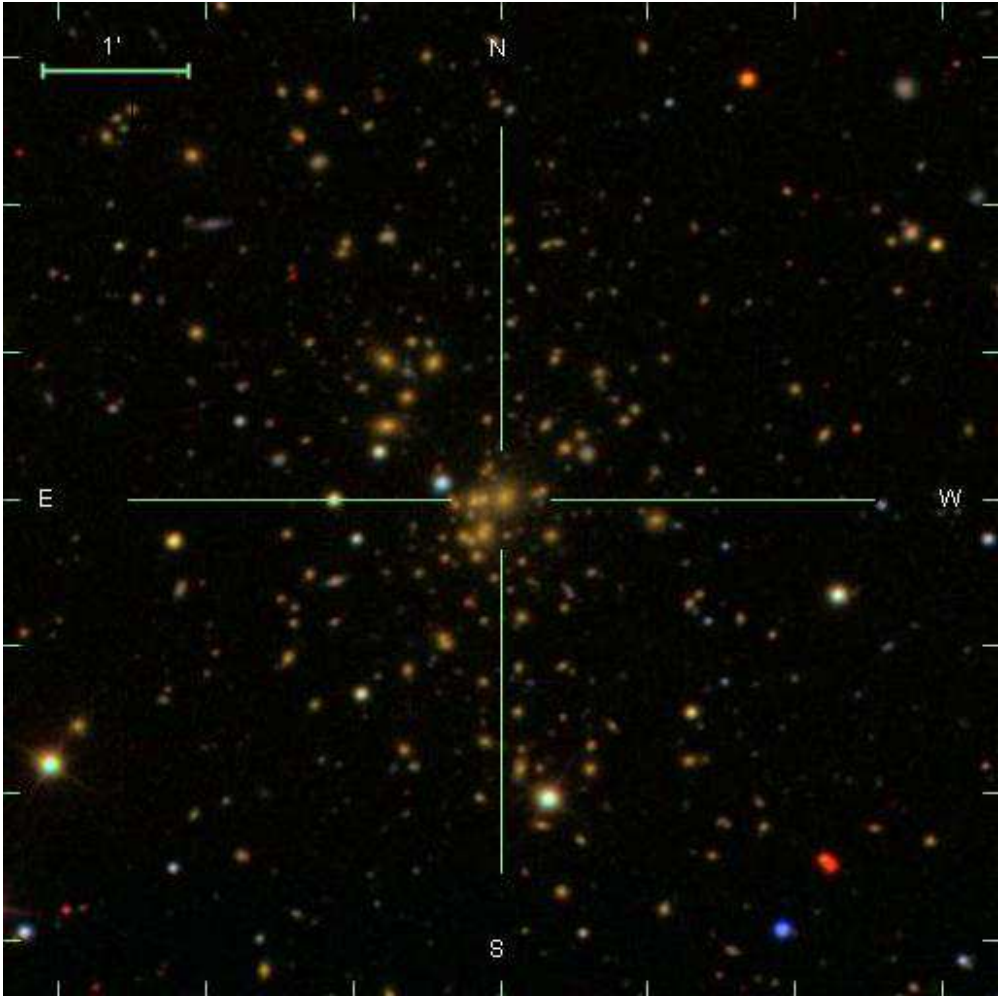}}}
\figcaption{Abell 1689, positioned on the maxBCG center. The FOV is
  $\simeq 0.75h^{-1}$ Mpc wide. The image is overwhelmed with luminous
  red cluster galaxies, essentially all at the same redshift.
  In \citet{abell89}, the position given is (ra,dec)=(197.8917,-1.365), which
  is 0.03 degrees different from the maxBCG postion, or $\simeq 230$ kpc. The
  redshift given by \citet{struble99} is $z=0.1832$, compared to $z=0.189$
  given by maxBCG. Also, note that the colors of these are less red than
  those from Figure 5, the $z=0.23$ cluster. See online edition of Journal for color image version.}
\label{fig:1689}.
\end{center}
\end{figure}

\begin{figure}
\begin{center}
\rotatebox{0}{\scalebox{0.8}{\plotone{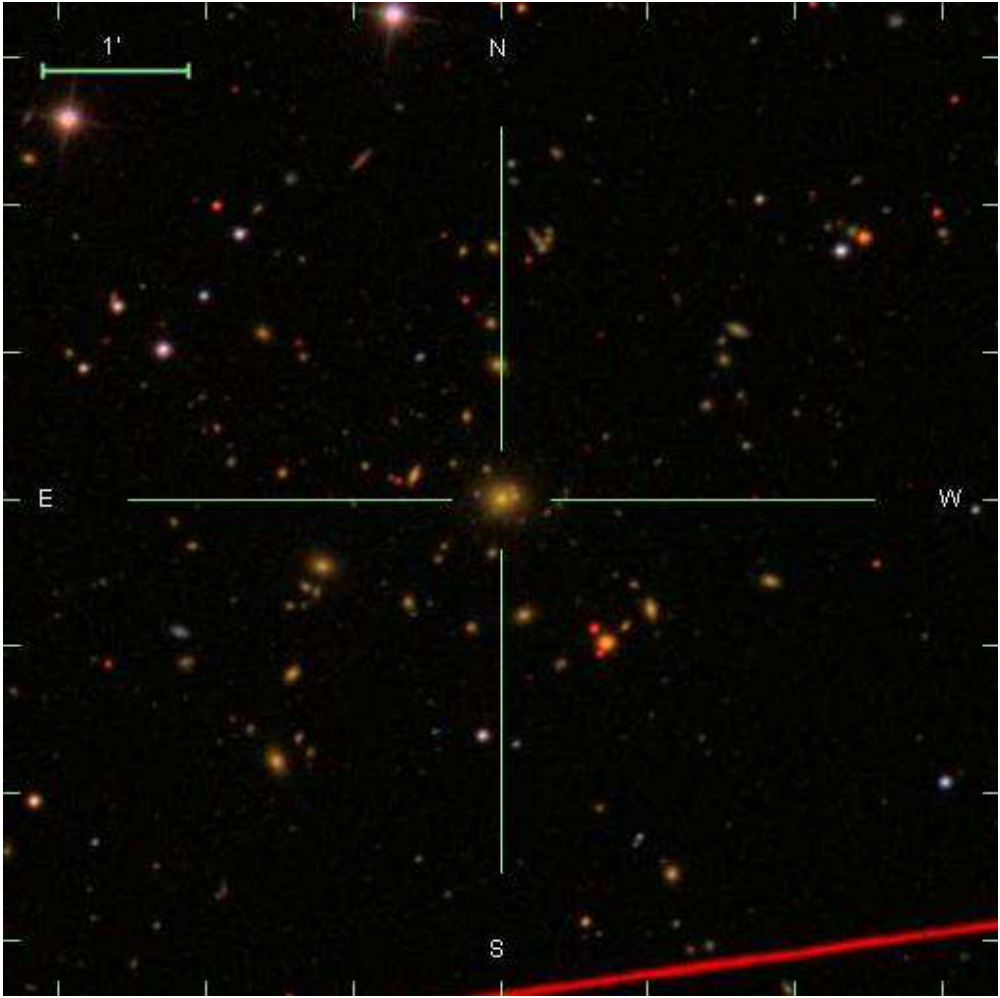}}}
\figcaption{NSC J131423-012734, at z=0.18 in the field of Abell 1689,
  $\sim 0.75$ degrees away. The FOV is also $\simeq 0.75h^{-1}$ Mpc,
  centered on the maxBCG position. The NSC catalog \citep{gal03} quotes
  a position at (ra,dec)=(198.5983,-1.4597), 0.016 degrees from the maxBCG
  center, or $\simeq 125$ kpc. The NSC photometric redshift of $z=0.247$ is not  consistent with maxBCG, which gives $z=0.181$ . SDSS spectroscopy provides
  redshifts for 4 cluster members, all at nearly $z=0.18$.
  See online edition of Journal  for color image.}
\end{center}
\end{figure}

\begin{figure}
\begin{center}
\rotatebox{0}{\scalebox{0.8}{\plotone{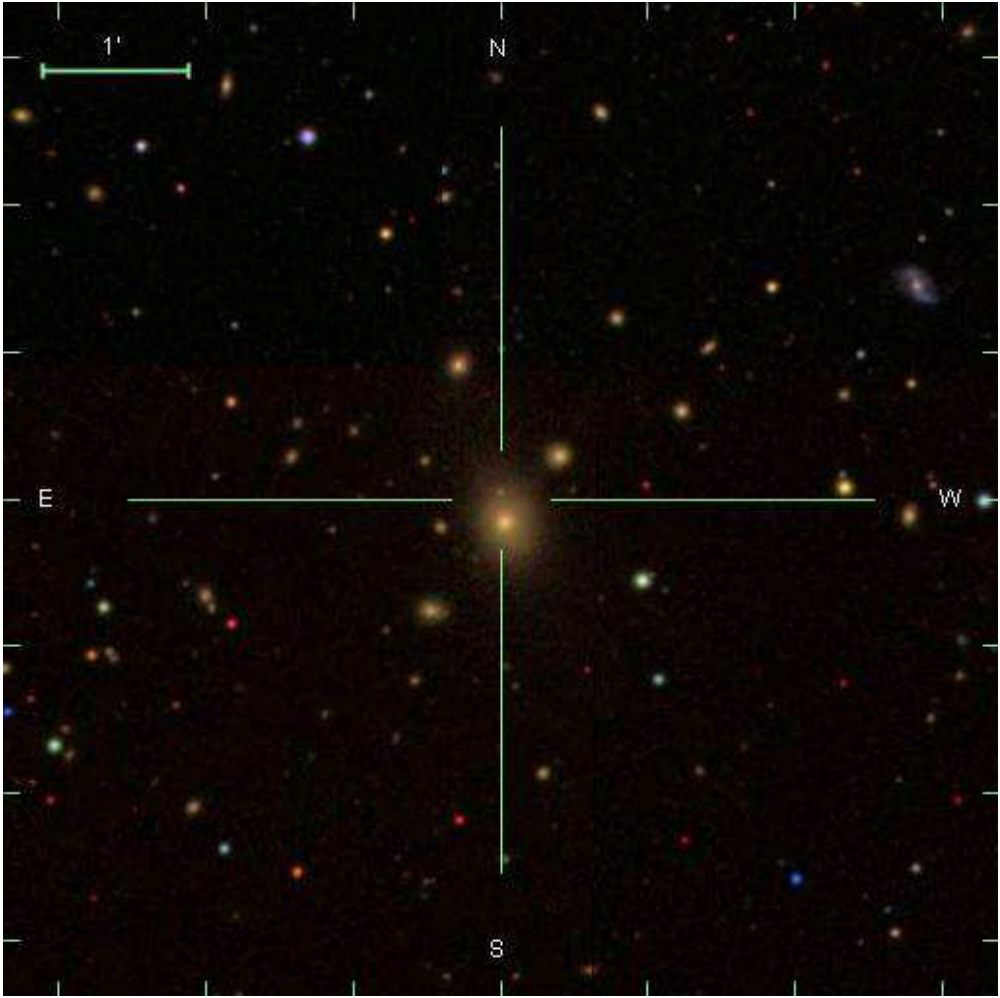}}}
\figcaption{MS 1306.7-0121, an X-ray selected cluster at $z=0.08$, in a
FOV spanning $\simeq 0.38h^{-1}$ Mpc. In \citet{stocke91}, the position is
given as (ra,dec)=(197.3254,-1.6228), 0.005 degrees from the maxBCG center, or $\simeq 20$ kpc. Its redshift is $z=0.088$, compared to the $z=0.08$ given by maxBCG. The BCG is clearly dominant
  and the colors of this lower redshift cluster are slightly less red
  than the previous two images, and quite noticeably bluer than
  Figure 5. See online edition of Journal for color image.}
\end{center}
\end{figure}

\begin{figure}
\rotatebox{90}{\scalebox{0.7}{\plotone{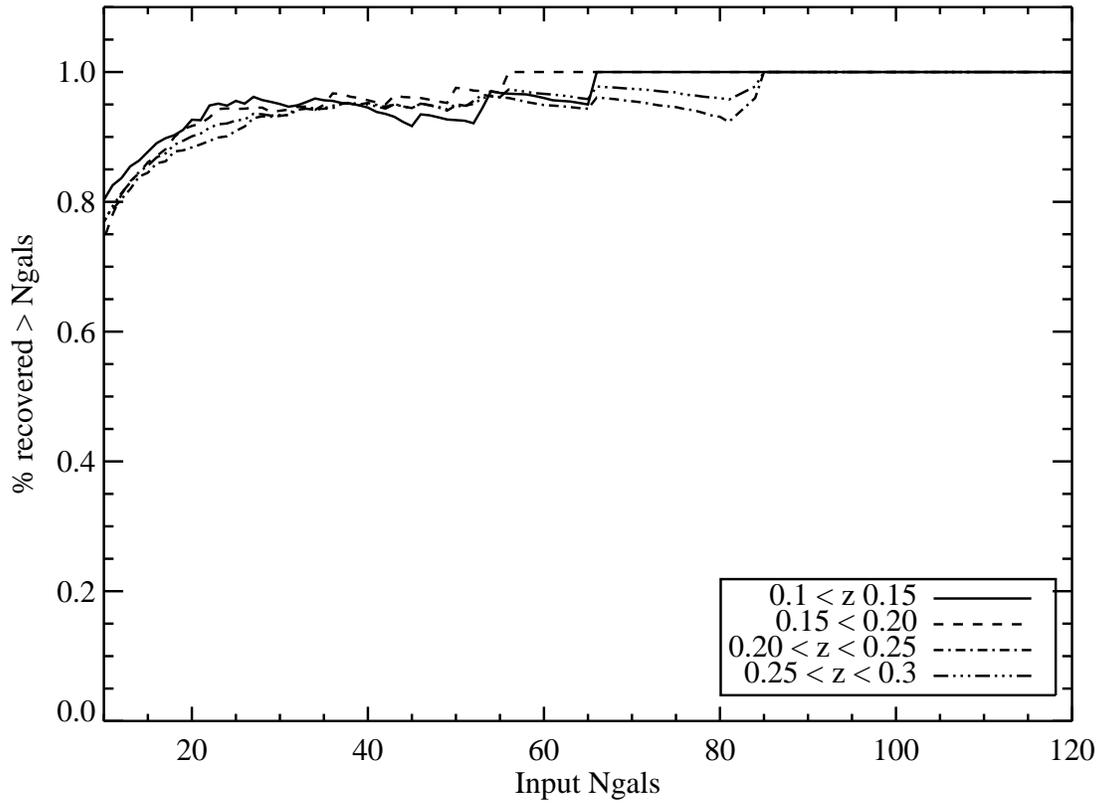}}}
\figcaption{Monte--Carlo Completeness of MaxBCG. The completeness is tested using shuffled catalogs with artificial clusters similar to \citet{goto02} (see text).}
\label{fig:mc_complete}
\end{figure}

\begin{figure}
\rotatebox{90}{\scalebox{0.7}{\plotone{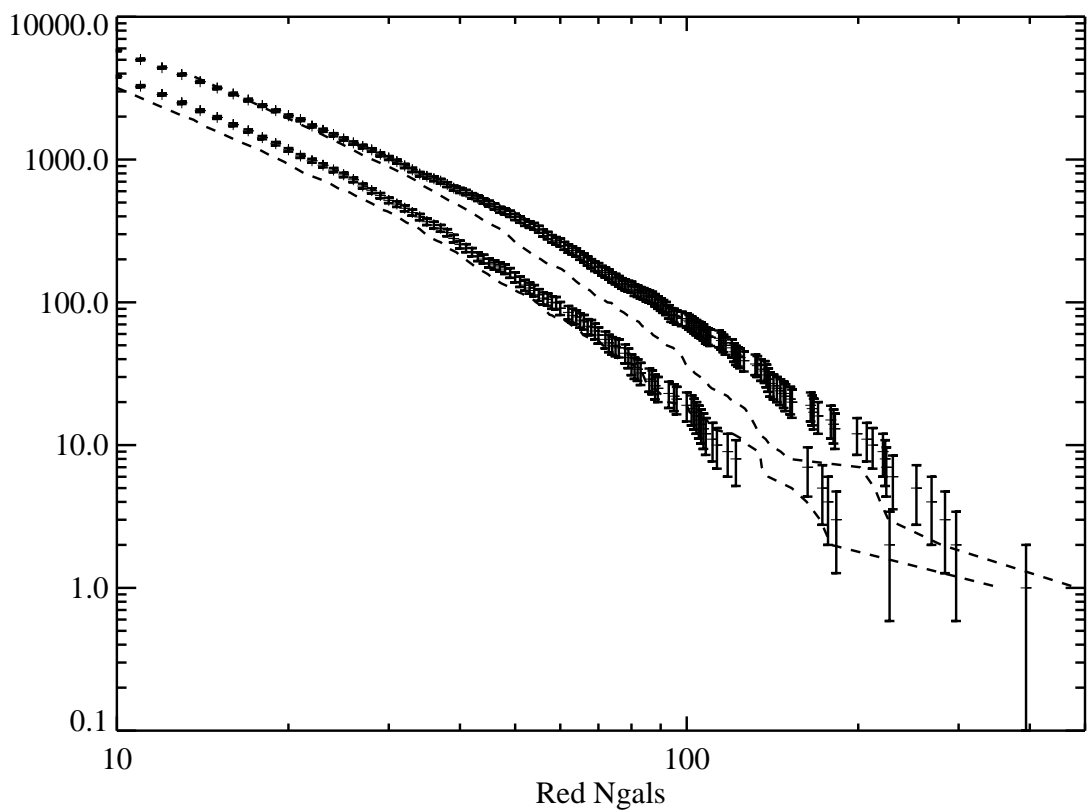}}}
\figcaption{Integrated counts for the derived cluster catalog (upper
  points, $N_{gals}^{r200}$) compared to the halo abundances, using
  $N_{int}^{red}$ as the halo richness. Poisson error bars are
  overplotted. The offset stems from richness estimates (\S 4.2) and
  fragmentation and overmerging (\S 4.3). In \S 4.2, a richness
  correction is determined by the simplest halo to cluster matching
  scheme. Assuming the differences are due only to richness
  estimation, the correction is applied to each abundance. The upper
  dotted line gives the new halo abundance after correction of the
  halo richnesses, and the new cluster abundance, where cluster
  richnesses are corrected, is given as the lower dotted
  line. Clearly, the process is not commutative.}
\end{figure}

\begin{figure}
\begin{center}
\rotatebox{0}{\scalebox{0.8}{\plotone{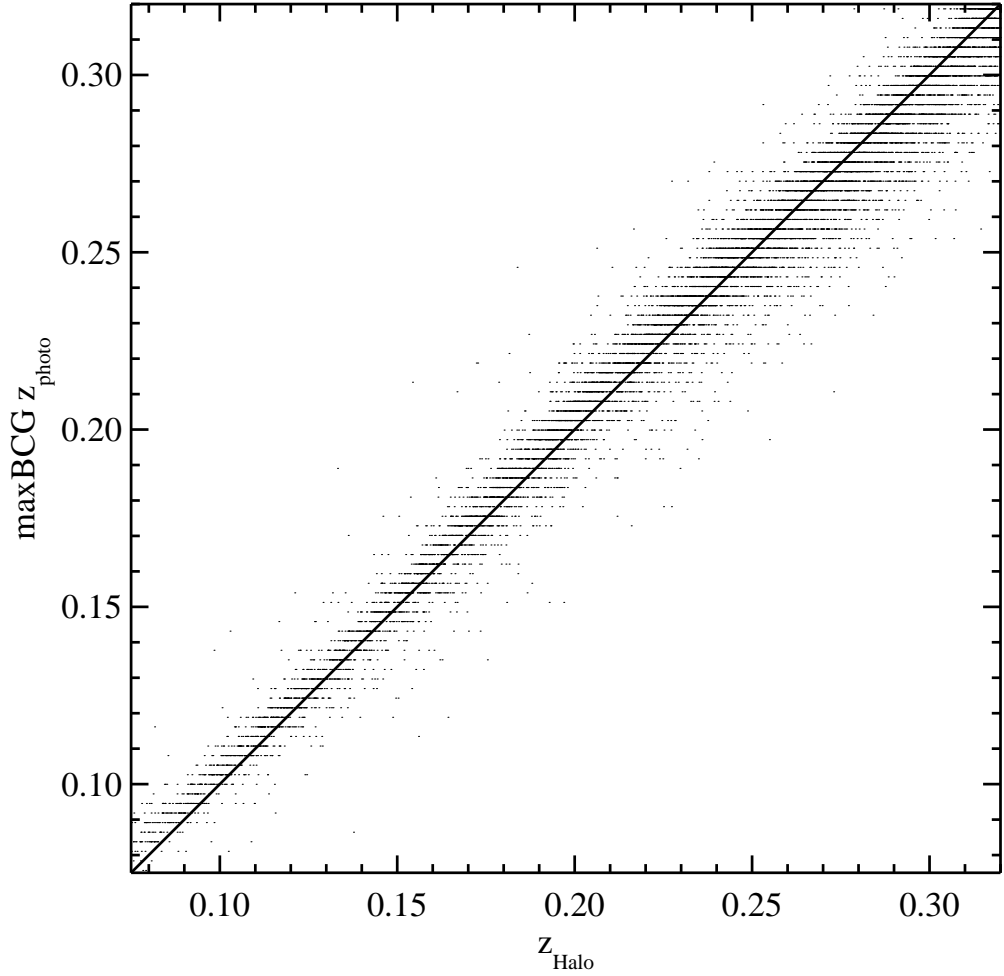}}}
\figcaption{Accuracy of the photometric redshift estimation for clusters.
  Plot compares the redshift identified for maxBCG clusters in the
  mock catalog with the redshift of the best-matched dark matter
  halo.}
\label{fig:zcomp}
\end{center}
\end{figure}

\begin{figure}
\begin{center}
\rotatebox{0}{\scalebox{0.7}{\plotone{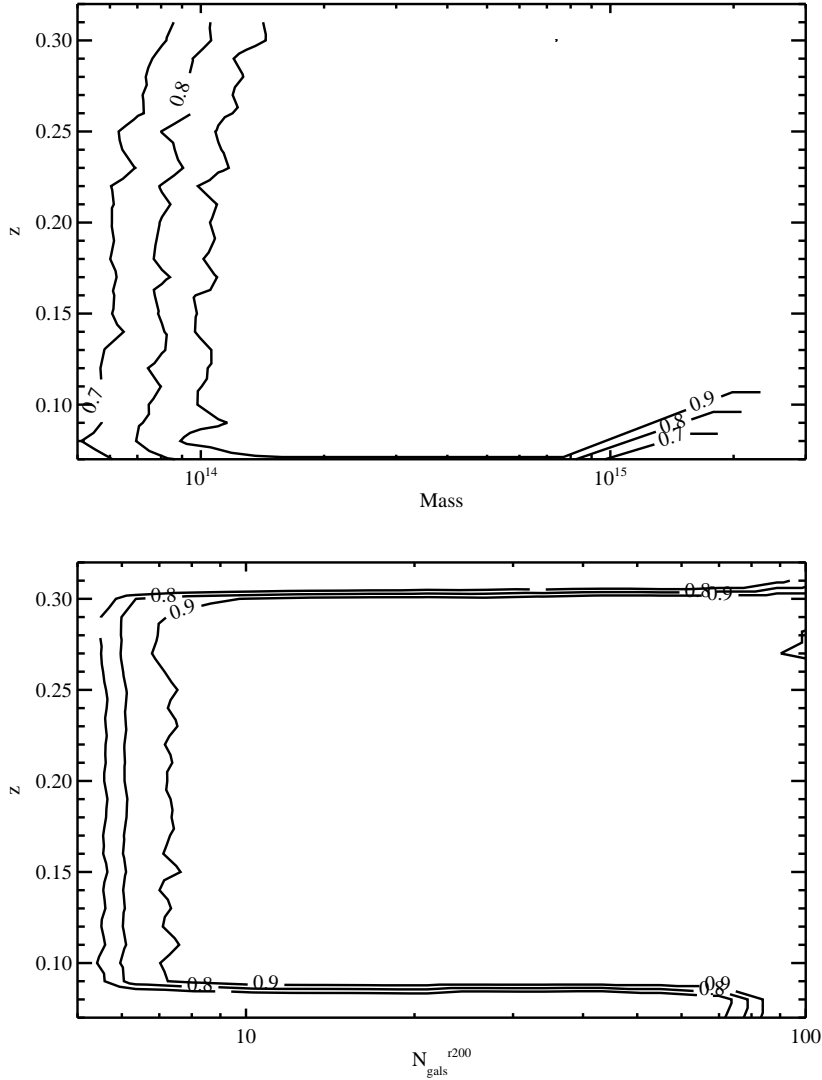}}}
\figcaption{Completeness and purity. The top panel shows contours of
  constant completeness in the redshift-mass plane, for a matching
  fraction of $f_h=0.3$. The lower panel shows contours of constant
  purity in the redshift $N_{gals}^{r200}-z$ plane, for a matching
  fraction of $f_c=0.3$. The decrease in completeness at low mass, and
  purity at low and high redshift are due to the chosen cuts on
  cluster richness and redshift: lower mass halos are found, but have
  $< 10$ red galaxies, and clusters at the redshift boundaries are
  often associated with halos just outside $z=0.1$ or $z=0.3$.}
\label{fig:completeness}
\end{center}
\end{figure}

\begin{figure}
\begin{center}
\rotatebox{0}{\scalebox{0.5}{\plotone{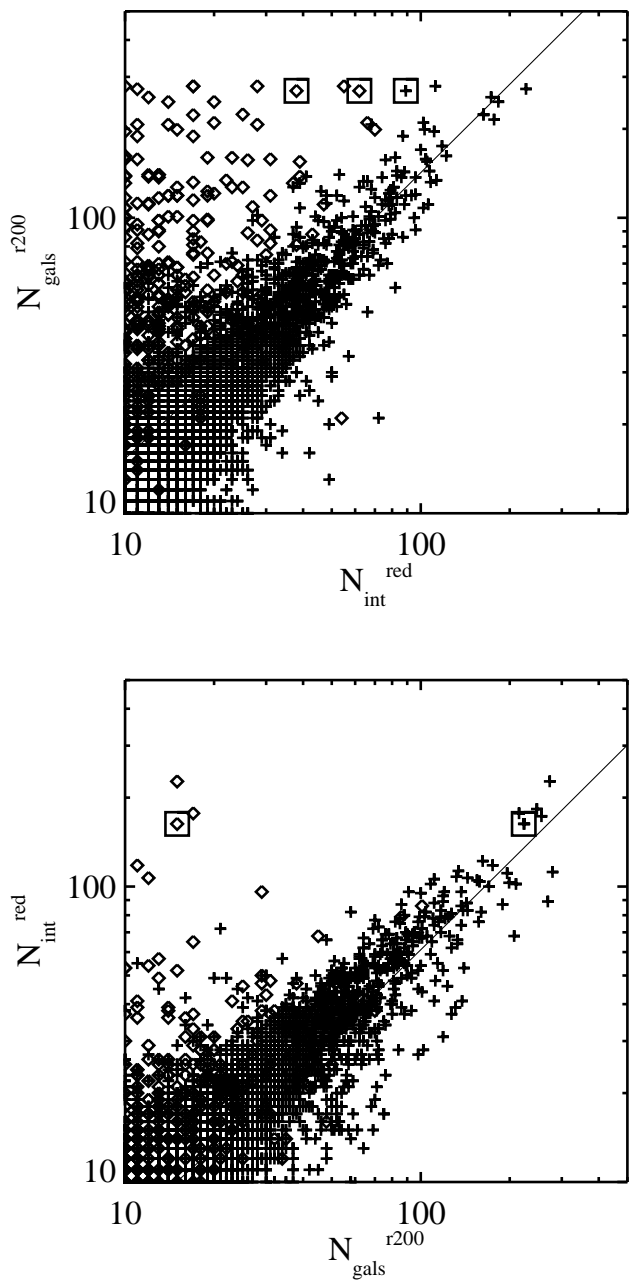}}}
\figcaption{Richness scalings between cluster and halo catalogs.
  $\emph{Top Panel}:$ Halos matched to clusters. For each halo
  $N_{gals}^{red}$ is plotted vs. $N_{gals}^{200}$ of the best-matched
  cluster. The diamonds represent duplicate halos (see text), and
  boxed points are examples cited in the text. $\emph{Bottom
    Panel}:$ Clusters are matched to halos. For each cluster
  $N_{gals}^{r200}$ is plotted vs. $N_{gals}^{red}$ of the
  best-matched halo. The diamonds represent duplicate clusters (see
  text).  Solid lines represent the mean relations in each case.}
\label{fig:nscaling}
\end{center}
\end{figure}

\begin{figure}
\begin{center}
\rotatebox{0}{\scalebox{0.7}{\plotone{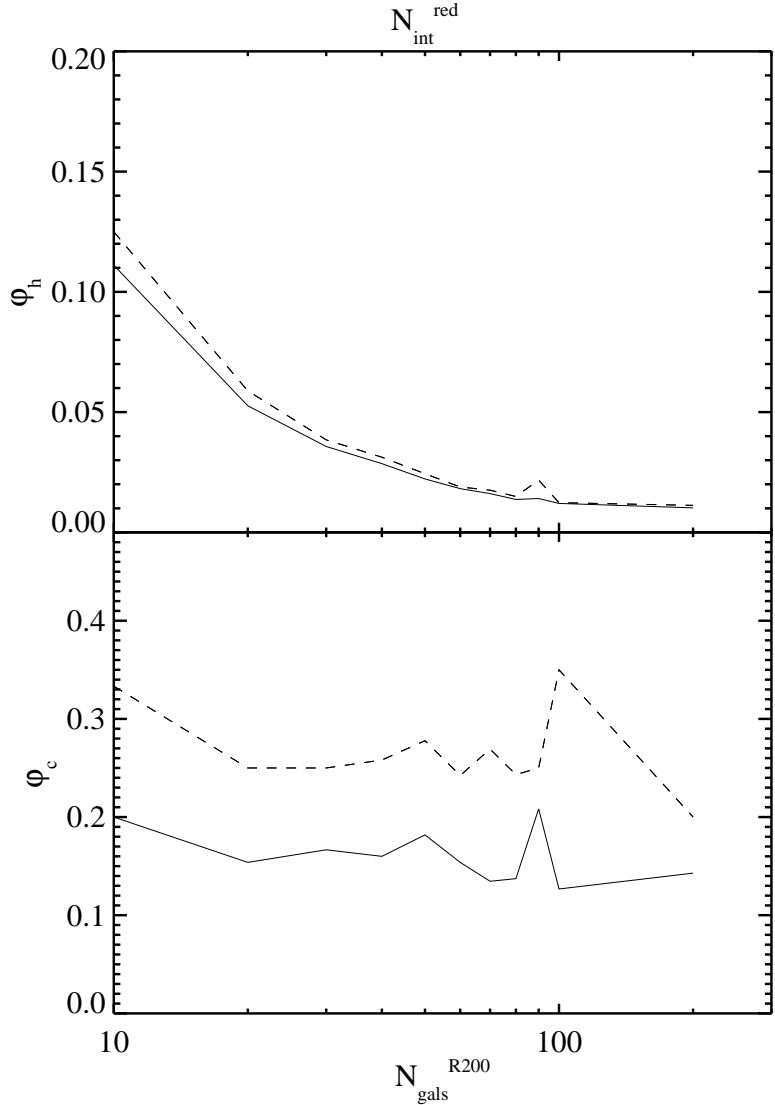}}}
\figcaption{Fragmentation and overmerging as a function of richness.
  Upper panel shows fragmentation of halos, as a function of their
  red-galaxy richness (as defined by Eq. \ref{eq:frag}); lower panel
  shows the overmerging of halos, as a function of cluster ridgeline
  richness.  The solid lines in upper and lower panels display the
  median amount of fragmentation and overmerging, respectively, while
  the dotted lines indicate the first quartile of the respective
  distribution. Fragmentation is nearly non-existent, while
  small-scale projection along the line of sight causes slight overmerging
  that boosts the cluster richness estimates by 10--20 percent.}
\label{fig:fragment}
\end{center}
\end{figure}

\begin{figure}
\rotatebox{0}{\scalebox{0.8}{\plotone{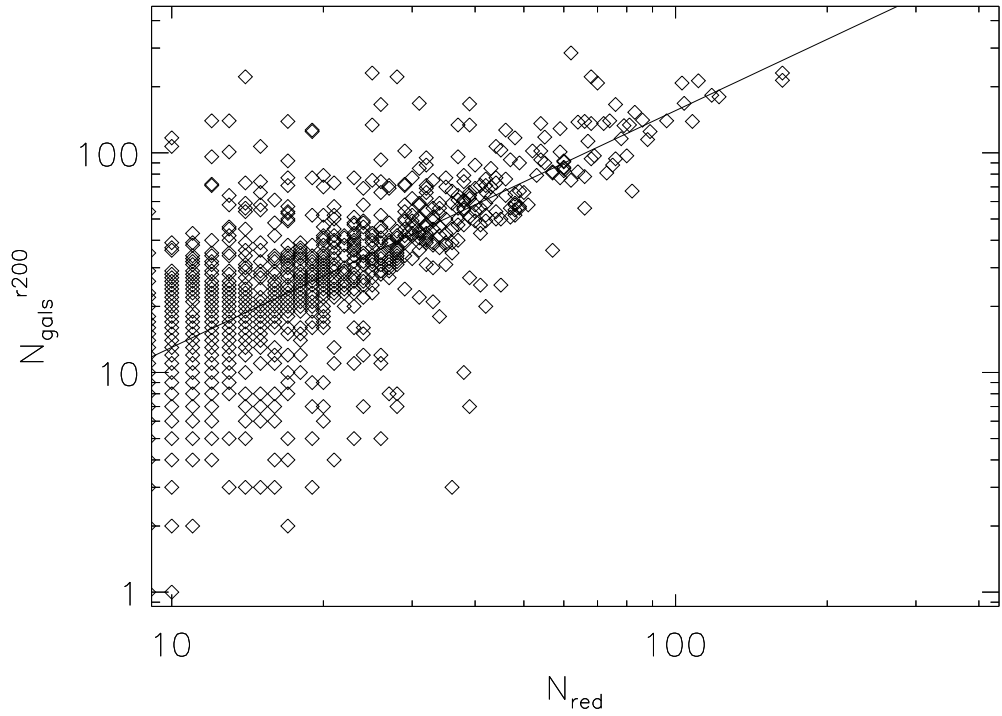}}}
\caption{Exclusive matching of clusters and halos, adapted from
  \citet{rozo06}. A full account of the algorithmic properties
  presented in section 4 enables a well-defined matching of halos and
  clusters, as presented in \citet{rozo06}. This is clearly a marked
  improvement on the upper panel of Figure 14, and
  demonstrates the efficacy of an exclusive matching algorithm.}
\label{fig:rozomatch}
\end{figure}

\end{document}